\documentclass[aip,rsi,preprint,graphicx]{revtex4-1} % for review purposes

\usepackage{graphicx}
\usepackage{units}
\usepackage{textcomp}
\usepackage{mathcomp}
\usepackage{amsfonts}

\draft % marks overfull lines with a black rule on the right

\begin{document}

%\preprint{}

\title{Variable Density Turbulence Tunnel Facility} %Title of paper



\author{E. Bodenschatz}
\email[]{eberhard.bodenschatz@ds.mpg.de}
%\thanks{}

\author{G. P. Bewley}
\email[]{gregory.bewley@ds.mpg.de}
%\thanks{}
%\affiliation{Max Planck Institute for Dynamics and Self-Organization, G\"ottingen, Germany}

\author{H. Nobach}
%\email[]{holger.nobach@ds.mpg.de}
%\thanks{}
%\affiliation{Max Planck Institute for Dynamics and Self-Organization, G\"ottingen, Germany}

\author{M. Sinhuber}
%\email[]{michael.sinhuber@ds.mpg.de}
%\thanks{}
%\affiliation{Max Planck Institute for Dynamics and Self-Organization, G\"ottingen, Germany}

\author{H. Xu}
%\email[]{haitao.xu@ds.mpg.de}
%\thanks{}
%\affiliation{Max Planck Institute for Dynamics and Self-Organization, G\"ottingen, Germany}

\noaffiliation
\affiliation{Max Planck Institute for Dynamics and Self-Organization, G\"ottingen, Germany}


\date{\today}

\begin{abstract}
The Variable Density Turbulence Tunnel (VDTT) 
at the Max Planck Institute for Dynamics and Self-Organization in G\"ottingen, Germany 
produces very high turbulence levels 
at moderate flow velocities, low power consumption and adjustable kinematic viscosity 
%change due to reviewer comment 1)
between $\unit[10^{-4}]{m^2/s}$ 
and $ \unit[ 10^{-7}]{m^2/s}$.  
The Reynolds number can be varied by changing the pressure or flow rate of the gas 
or by using different non-flammable gases including air.  
The highest kinematic viscosities, and hence lowest Reynolds numbers, are reached with  air or nitrogen at 0.1\,bar.  To reach the highest Reynolds numbers the tunnel is pressurized to 15\,bar 
with the dense gas sulfur hexafluoride (SF$_6$).  
%change due to reviewer comment 1)

Turbulence is generated at the upstream ends of two measurement sections with grids, 
and the evolution of this turbulence is observed as it moves down the length of the sections.  
We describe the instrumentation presently in operation, 
which consists of the tunnel itself, classical grid turbulence generators, 
and state-of-the-art nano-fabricated hot-wire anemometers 
provided by Princeton University.\cite{vallikivi:2011}  
We report measurements of the characteristic scales of the flow 
and of turbulent spectra 
up to Taylor Reynolds number $R_\lambda \approx 1600$, 
higher than any other grid-turbulence experiment.  
We also describe instrumentation under development, 
which includes an active grid 
and a Lagrangian particle tracking system that moves down the length of the tunnel with the mean flow.  
In this configuration, 
the properties of the turbulence are adjustable and its structure is resolvable 
up to $R_\lambda \approx 8000$.  
\end{abstract}


\maketitle %\maketitle must follow title, authors, abstract and \pacs

% References should be done using the \cite and \label commands
\section{Introduction}
%\label{}

% If in two-column mode, this environment will change to single-column format so that long equations can be displayed. 
%\begin{widetext}
%$$\mbox{put long equation here}$$
%\end{widetext}

% Figures should be put into the text as floats. 
% Use the graphics or graphicx packages (distributed with LaTeX2e). EPSFig is no longer fully supported.
% Put the label that you will use with \ref{} command in the braces of the \label{} command.
%
% \begin{figure}
% \includegraphics{}% % Important NOTE: Please make certain your figures do not include local directory paths. ex. "c:\file\sub\fig1.eps"
% \end{figure}

% \begin{table}
% \end{tabular}
% \end{table}

\subsection{On the need for the VDTT}

Turbulence plays a decisive role in the universe. 
Its influence extends broadly throughout nature and technology.\cite{corrsin:1961}  
For example, turbulence controls the spread of trace gases,\cite{roth:2000,arnfield:2003} 
pollutants,\cite{james:2002} and particulate matter \cite{shaw:2003} in the atmosphere and oceans, 
the mixing of fuel and air in engines,\cite{pope:2013} 
the generation of energy-draining wakes behind airplanes and cars,\cite{bechert:2000} 
and even the evolution of planets, stars and the universe as a whole.\cite{gibson:1996,ryu:2008}  

What underlies all turbulent motion is the balance between 
the inertia of the fluid 
and the pressure and friction forces that the fluid exerts on itself.  
In almost every practical setting 
this balance is complicated by additional effects, 
such as 
buoyancy-driven convection where 
temperature gradients drive the flow, 
\cite{ahlers:2009} 
rotation-induced 
Coriolis accelerations in 
oceanic, atmospheric flows on earth or other planets, 
\cite{lindborg:2005} 
electromagnetic forces 
in conducting fluids like those that make up the sun,\cite{krommes:2002} 
nonlinear stresses in non-Newtonian fluids like blood,\cite{antiga:2009} 
or changes in material properties in flames.\cite{peters:2000}  

If we want to discover something generic about turbulence, 
something that is essential wherever turbulence is fundamental, 
we may limit our inquiries to its most essential ingredients: 
inertia, pressure, and friction.  
Such turbulence can be created 
by mechanically stirring a liquid or gas.  
In this spirit of interest in universality, 
we may also wish to exclude the influences of the geometry of the flow.  
We then want to study a flow that minimizes the effect of the boundaries of its container, 
and does not exhibit a preferred orientation.\cite{biferale:2005}  
Such a flow is called statistically homogeneous and isotropic,\cite{taylor:1935} 
and a close approximation of it can be realized in a wind tunnel 
by disturbing a uniform free-stream flow with a mesh or grid.\cite{corrsin:1961,comte-bellot:1966}  

When turbulence is well-developed, 
it comprises large sweeping motions that contain most of the mechanical energy of the flow, 
and relatively sharp gradients that dissipate most of this energy.\cite{frisch:1996}  
The Reynolds number, $Re = \rho u^{\prime} L / \mu$, 
gauges the separation between these large and small scales.  
In other words, high Reynolds numbers mean large scale separations.  
Here, $\rho$ and  $\mu$ are the density and dynamic viscosity of the fluid, respectively, 
$u^{\prime}$ is the amplitude of its velocity fluctuations, 
and $L$ a characteristic scale over which motion is correlated.  
For typical values in the atmosphere, for example, 
the Reynolds number may reach values of a million, or more.  

Universal features of turbulence are expected to reveal themselves at high Reynolds numbers.  
That is, when the scale separation between the energetic and dissipative motions is large, 
and where at sufficiently small scales the flow may forget its initial conditions.  
Experiments at high Reynolds numbers are desirable, then, 
not only to gain insight into natural flows, 
but to provide clues about a fundamental description of turbulence.  
The question that remains is how to produce well-controlled, high-Reynolds-number turbulence 
whose properties are commensurate with current measurement technology.  

The constraints in designing a facility to generate high-Reynolds-number turbulence 
are that it is realizable within the available space, time and funds, 
and that the turbulence it produces is observable with available technology.  
In this context, 
each of the density, viscosity, velocity and length scales must be optimized 
in order to maximize the Reynolds number.  
The fluctuating velocity and length scales $u^{\prime}$ and $L$ 
determine the size of a facility and its power consumption.  
The fluctuating velocity, $u^{\prime}$, is typically a fixed fraction of the mean flow speed, $U$, 
and the power required to run a facility scales with $U^3$.  
Clearly, modest flow velocities and a modest size 
minimize both construction and operational costs.  
We are left to find 
a fluid with high density and low viscosity.  

The VDTT uses pressurized gases to satisfy the requirement that the fluid density be high 
while the viscosity is low.  
Gases have low dynamic viscosities, $\mu$, which depend only weakly on pressure.  
Thus the kinematic viscosity, $\nu = \mu/\rho$, 
can be decreased simply by increasing the pressure and thus the density $\rho$ of the gas.  
The VDTT is equipped to use any non combustible gas, 
and Sulfur Hexafluoride (SF$_6$) in particular.  
SF$_6$ has the advantage that at atmospheric pressure it is 5 times denser than air, 
and at a pressure of 15 bar it reaches a density of about one tenth of water.  
This pressure is much lower than the 100 bar that would be needed to bring air to the same density.  
As a consequence, pressure vessels can be limited to 15\,bar and can be built thin-walled 
and economically.  
By this method, the Reynolds number can not only be high, 
but can also be regulated over more than two decades by changing pressure alone.  
That is, without changing the conditions of the flow at the large scale; % {\textit i.e.},   
this translates to flexibility in controlling the experimental conditions.  
In addition, SF$_6$ has the advantage of not being hazardous to life 
other than by replacing air and the oxygen it contains.  
It has the disadvantage, however, of being a greenhouse gas; 
great care needs to be taken not to lose it to the environment.  

The VDTT is a wind tunnel 
around which pressurized gases circulate in an upright, closed loop.  
At the upstream ends of two test sections in the VDTT, 
the free stream is disturbed mechanically.  
The resulting turbulence evolves down the length of the tunnel without 
the middle region being substantially influenced by the walls of the tunnel.  
The test sections are long enough (8\,m) that the turbulence evolves 
through at least one eddy turnover time, $L/u^{\prime} \approx 1\,s$, 
during its passage down the tunnel.  
That is, the turbulence can be observed over the time it takes energy to cascade 
from the large scales all the way down to the dissipative ones.  

The width of the cross section of the tunnel constrains the characteristic scale $L$ of the flow, 
and for fixed $L$ higher Reynolds numbers lead to ever smaller scales of motion.  
In the VDTT, the cross section is wide enough (1.5\,m) that even at the highest Reynolds numbers, 
the smallest scales of motions are neither too small nor too fast to be resolved by existing 
state-of-the-art instrumentation.  
We note that scales of a modest size are also desirable if one wants to use optical techniques 
to resolve all scales of the flow.  
With current technologies large optical components, like lenses, mirrors and detectors, 
are difficult to produce.  

The VDTT was designed around a Lagrangian particle tracking system (LPT).  
This requirement implies a limit on the mean flow rate $U$.  
The VDTT's maximum flow speed of $U = 5\,m/s$ is sufficiently small that we can accelerate 
the LPT system to match the flow speed at the upstream end of a test section, 
move it at mean-flow speed down the tunnel, 
and stop it at the downstream end.  
We can then follow the evolution of the flow in the frame of the flow itself, 
by imaging the motions of particles carried by the flow.  
Eulerian measurements are not excluded, and the data we present here 
were acquired with both traditional hot wires and the new NSTAP probes.\cite{Bailey2010,vallikivi:2011}  
These were the constraints on the design of the VDTT.  

The history of using pressurized gases goes back almost 100 years.  
We provide in the remainder of this introduction 
a short historical review of predecessors to the VDTT.  
That is, of wind tunnels in which the density of the working fluid could be varied.  
There are several other types of wind tunnels and a great variety of turbulence facilities in general 
that we do not review, 
and which are worthy of an article in and of themselves.  
We continue in Section II with a description of the apparatus and its technical details.  
In section III, we describe velocity measurements and their statistics, 
and we conclude with an outlook in Section IV.

\subsection{Historical Review of Variable Density Facilities}

Variable density wind tunnels have been an important tool for aeronautical research and development 
for almost 100 years.  
Since both the density and the speed of the wind in such tunnels 
could be adjusted independently, 
both the Reynolds number and the Mach number of the flows could be set independently.  
This made it possible with small-scale models to observe the aerodynamics 
of full-scale airplanes 
under well-defined laboratory conditions.  
Before the advent of computers, 
these tunnels provided the \emph{only} way to test design ideas.  
We emphasize that even with today's computers, 
wind tunnel tests remain essential in the development of airplanes.  
In the following, we give the maximum Reynolds numbers attainable in the various facilities 
in terms of its mean flow speed and 10\% of its width, so that $Re_{WT} \equiv 0.1\sqrt{A}U/\nu$, 
where $A$ is the cross sectional area of the tunnel.  
The VDTT reaches values of this Reynolds number up to $Re_{WT} = 4.4\cdot10^6$, 
among the highest yet achieved despite the low speed and long test section of the tunnel.  

The first wind tunnel in which the density of the working fluid could be adjusted 
was the 
``Variable Density Wind Tunnel of the National Advisory Committee for Aeronautics'' 
(the VDT, with one ``T'' instead of two).  
It was designed by Max Munk,\cite{Munk1921} 
a student of Ludwig Prandtl.  
Prandtl incidentally founded the Kaiser Willhelm Institute for Flow Research 
in G\"ottingen, Germany.  
This institute is the predecessor of the MPI for Dynamics and Self-Organization, 
which is the residence for the new VDTT, the subject of this article.  
The original VDT was built at the Langley Research Center in Virginia 
and became operational in 1923.  
As must be the case in all pressurized wind tunnels, 
the gas in the tunnel went around a closed circuit 
rather than being open to draw air in from the surrounding environment.  
Recirculating tunnels such as the VDT 
are called G\"{o}ttingen-type tunnels.\cite{Oswatitsch1987}  
The test section of the VDT 
was housed in a 10.2\,m long, 4.6\,m wide cylindrical pressure vessel 
that could withstand pressures up to 21\,bar.\cite{Munk1926}  
The tunnel was used to test airfoils and model airplanes in states 
corresponding to various atmospheric conditions.  
The original tunnel was made of wood, and was destroyed by a fire in 1927.  
It also faced serious difficulties with vibration and flow quality, 
so that the tunnel was redesigned 
as a more rigid and fire-proof structure 
when it was rebuilt in 1930.\cite{Jacobs1933}  

The VDT produced a huge amount of airfoil data, 
reaching Reynolds numbers of about $Re_{WT} = 5.4\cdot10^6$.  
But high free-stream turbulence intensities led to inaccurate measurements of drag.  
To overcome this problem, 
the ``Langley two-dimensional low-turbulence pressure tunnel'' (LTPT) 
was designed and completed in 1938, 
also at the Langley Research Center.\cite{Doenhoff1947}  
The LTPT was large, at 44.5\,m long and 17.7\,m wide.  
By design, the airfoils in this tunnel could span the whole test section, 
effectively reducing by one the dimension of the flow.  
The tunnel ran at pressures up to 10\,bar 
and at Reynolds numbers of up to about $Re_{WT} = 6.1\cdot10^6$, 
A low turbulence level of less than 1\% was achieved through a contraction 
with a large area ratio 
in combination with a series of wire screens.  
This tunnel was operational for many years 
and underwent massive modifications in the early 1980's.\cite{McGhee1984}  
It was still in operation during the early $21^{st}$ century \cite{Choudhari2002} 
until the drive motor burned in 2006.  
Demolition of the tunnel began after no further funding for repair was granted.  
Further information can be found on the NASA website.\cite{WebsiteNasa}  

Several years after the construction of the original VDT, 
interest in variable density tunnels spread over to Europe.  
This can be seen in the construction of the ``Compressed Air Tunnel'' 
at the National Physical Laboratory in Teddington in 1931.\cite{Pankhurst1972}  
In its 6\,foot diameter test section, 
air could be compressed up to 25\,bar to produce a wide range of Reynolds numbers 
up to $Re_{WT} = 8\cdot10^6$.  
As with the VDT, the main focus of the Compressed Air Tunnel 
was the aerodynamical improvement of aircraft.  

A different approach to model testing was chosen in Braunschweig 
with the construction of the 
``Variable density high speed cascade wind tunnel'' 
at the Deutsche Forschungsanstalt f\"{u}r Luftfahrt 
in 1956.  
In this tunnel, the pressure could be lowered from 1 to 0.1\,bar 
to test the performance of blade cascades under various conditions.  
The Reynolds and Mach numbers in the tunnel could be set independently,\cite{Schlichting1956} 
up to a maximum Mach number of 1.1 
and Reynolds number $Re_{WT} = 4\cdot10^6$.  

While there were several experiments focused on airfoils or model planes 
in variable density wind tunnels, 
relatively few exist that addressed fundamental turbulence questions.  
The first was probably the study of Kistler and Vrebalovich, 
which reports on classical grid turbulence experiments at the 
``Southern California Co-operative Wind Tunnel'' \cite{Millikan1948} 
just before it closed in the 1960's.\cite{Kistler1966}  
The measurements were made at wind speeds of up to 60\,m/s 
and with air pressures between 0.2\,bar and 4\,bar.  
The size of the cross section was 2.6\,m $\times$ 3.5\,m, 
so that Reynolds numbers up to $Re_{WT} = 4.5\cdot10^6$ were possible.  
They observed turbulence produced by classical grids, 
so that the grid Reynolds number $Re_M = \rho U M / \mu$ is relevant, 
where $M$ is the mesh spacing of the grid.  
The combination of high pressure, high velocities, and large cross section  
allowed for measurements of both the decay and the spectrum 
of grid turbulence at grid Reynolds numbers up to 2.4 million, 
higher than any previous study, or indeed than any subsequent study.  
The grid turbulence data we introduce in this paper are the first to exceed in Reynolds number 
those of Kistler and Vrebalovich; we reach 4.9 million.  

A second series of fundamental studies in pressurized tunnels 
was conducted at the Nuclear Research Laboratories in J\"{u}lich in the 1970's.  
The wind tunnel there could be filled with Helium of up to 40\,bar.  
While the main focus of the facility was heat transfer in heat exchangers for power plants, 
some work was done on the flow past spheres.\cite{Achenbach1972}  
Here, the Reynolds numbers reached $Re_{WT} = 3.1\cdot10^5$.  

A relatively recent facility was the ``high pressure wind tunnel'' (HDG) 
at the German Aerospace Center (DLR) in G\"{o}ttingen.\cite{Foersching1981}  
The HDG withstood pressures up to 100\,bar 
and ran at speeds of up to 35\,m/s, 
so that Reynolds numbers up to $Re_{WT} = 1.3\cdot10^7$ were possible.  
It has seen extensive use in studies of models and of 
physical effects 
such as flow-induced resonances.\cite{Schewe1989}  
For example, 
the classical problem of flow around a circular cylinder was investigated 
in the range $10^4 < Re < 10^7$ merely by varying the flow parameters 
and not the size of the cylinder.\cite{schewe:1983,schewe:2001}  

In the late 1990's, 
the Princeton/DARPA-ONR SuperPipe Facility was built at the Princeton Gas Dynamics Lab Facilities.  
It was a 34\,m long and 1.5\,m wide pipe flow facility with a 12.9\,cm test section diameter.  
It circulated air pressurized to between 1 and 220\,bar,\cite{Zagarola1997} 
so that Reynolds numbers reached $Re_{WT} = 2.3\cdot10^6$.  
This highly pressurized air made it possible 
to uncover the scaling laws of pipe flows at very high Reynolds numbers, 
$\mathrm{Re^+}= Ru_\tau/\nu$, 
of about $10^5$.  
Here $R$ is the radius of the pipe and $\nu/u_\tau$ the viscous length scale.\cite{Hultmark2012}  

To allow for model testing, a grant for a wind tunnel based upon the SuperPipe Facility 
was granted in 1998 and led to the construction of 
the Princeton/ONR High Reynolds Number Testing Facility.  
This wind tunnel operated at up to 240\,bar.  
With two 2.6\,m test sections of 46\,cm diameter, 
it allowed for submarine model testing at high Reynolds numbers up to $Re_{WT} = 9.6\cdot10^6$.
A more detailed description of this and the SuperPipe facility 
can be found on the website of the Princeton Gas Dynamics Lab Facilities.\cite{WebsitePrinceton}  

\section{Construction Details}

%\subsection{Overview of the facility}

% the tunnel - the pressure vessel
% the working fluids
% the gas handling system
% the safety system
% the fan
% the heat exchanger
% the filter bypass 
% the measurement sections
% electrical and optical access
% the linear traverse
% the grids
% the measurement systems

The facility at the Max Planck Institute in G\"ottingen 
comprises a pressure vessel, 
the associated gas handling apparatus, 
safety systems, 
grids to generate turbulence, 
and turbulence measurement systems.  
We describe these systems in some detail in the following sections.  
We also describe some of the measures undertaken to control the 
homogeneity of the flow in the measurement sections in Sec.~\ref{sec:aero}.

\begin{figure}
\centerline{\scalebox{8}{\includegraphics{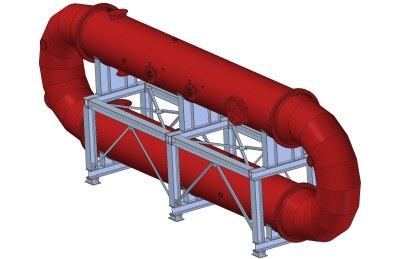}}}
%\centerline{\scalebox{0.36}{\includegraphics{pictures/DSC_0145.jpg}}}
\centerline{\scalebox{0.96}{\includegraphics{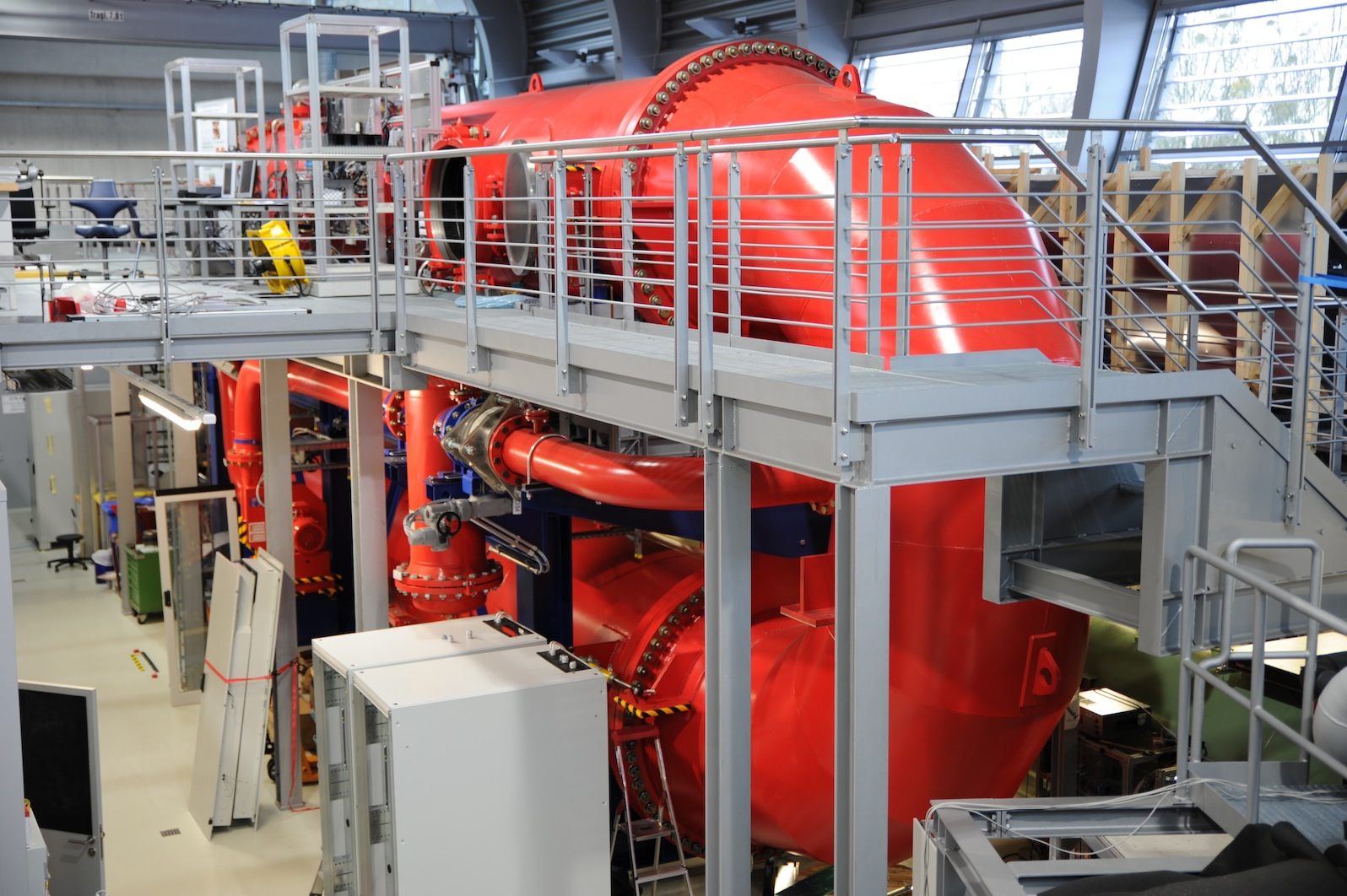}}}
\caption{3D view and photograph of the Variable Density Turbulence Tunnel.  }
\label{fig:tunnel}
\end{figure}

\subsection{Pressure Vessel}
\label{sec:vessel}

\begin{figure}
\centerline{\scalebox{0.5}{\includegraphics{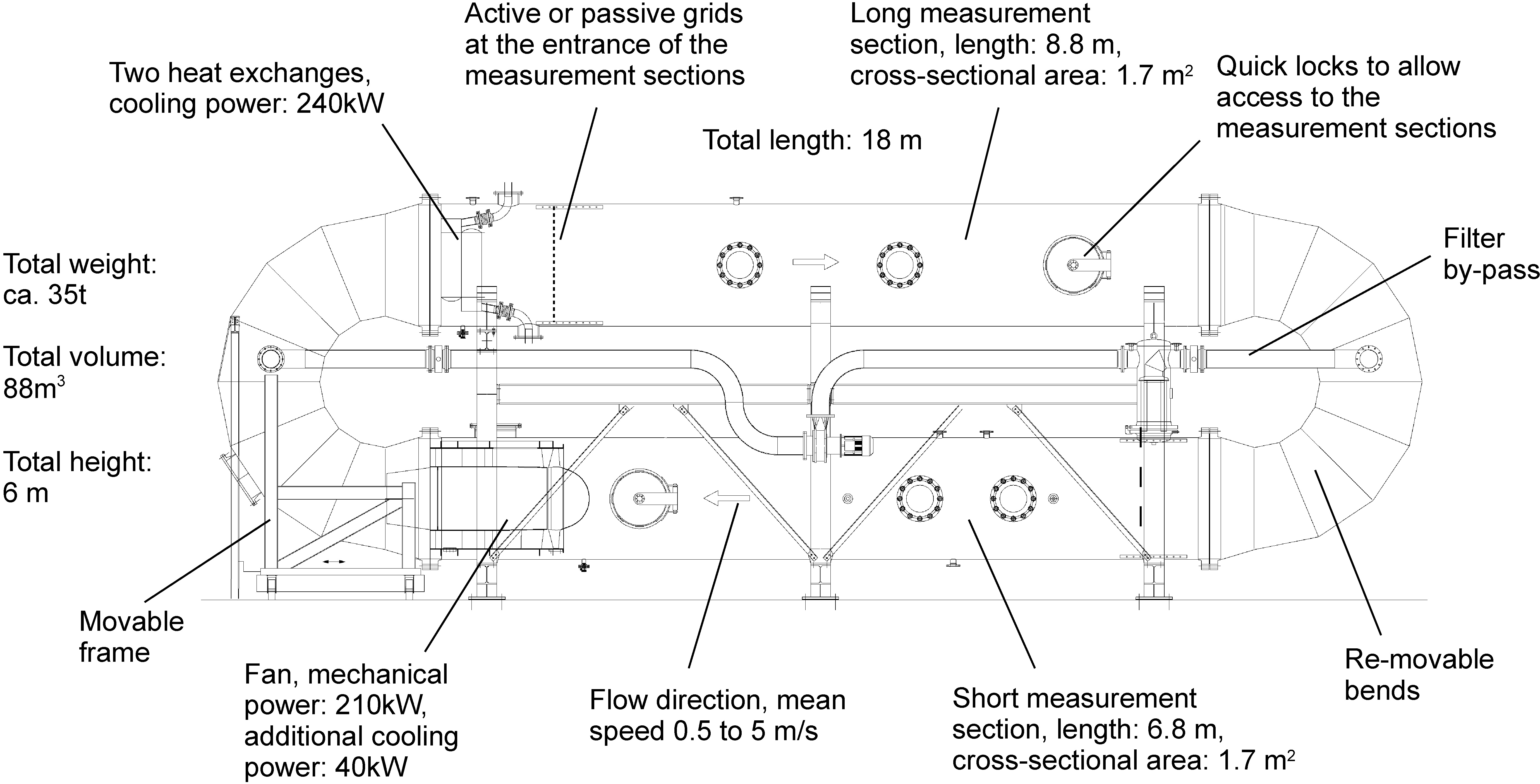}}}
\caption{Sketch of the VDTT.}
\label{fig:tunnel_sketch}
\end{figure}

\begin{figure}
\centerline{\scalebox{0.96}{\includegraphics{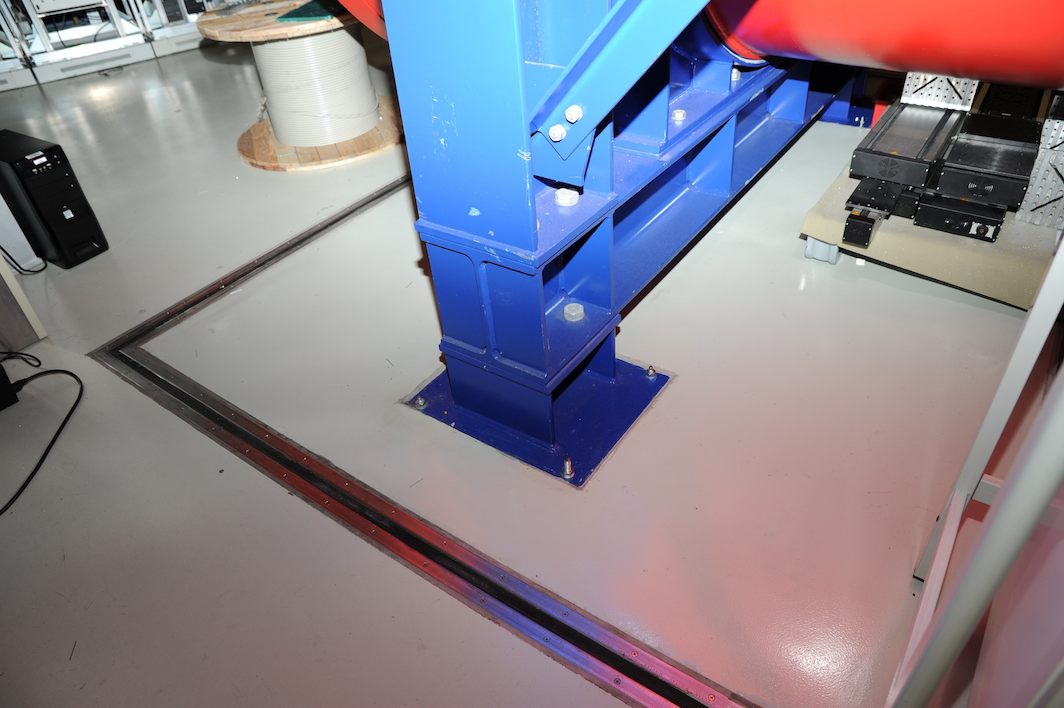}}}
\caption{Vibration isolating foundation.  
There is one of these under each of the three supports on which the VDTT rests.  }
\label{fig:fundament}
\end{figure}

As can be seen in Figs.~\ref{fig:tunnel} and \ref{fig:tunnel_sketch}, 
the pressure vessel is upright, $\unit[18.2]{m}$ long, and $\unit[5.3]{m}$ high.  
It consists of two $\unit[11.68]{m}$ long straight cylindrical sections 
with an inner diameter of $\unit[1.84]{m}$, 
and two elbows with a center-line 
radius of $\unit[1.75]{m}$ and an inner diameter of $\unit[1.52]{m}$.  
The total volume of the pressure vessel is $\unit[88]{m^3}$, 
which corresponds to $\unit[8.7]{t}$ of SF$_6$ 
at $\unit[15]{bar}$ and $\unit[20]{\tccelsius}$.  
The pressure vessel is made of P265GH (1.0425) steel, 
the straight tubes are $\unit[20]{mm}$ thick, and two elbows are $\unit[18]{mm}$ thick.  
The support frame is made of S 235 JRG2 steel.  
To reduce mechanical coupling between the tunnel and the building, 
the tunnel rests on three vibration isolated foundations (Fig.~\ref{fig:fundament}).  

\begin{figure}
\centerline{\scalebox{1.12}{\includegraphics{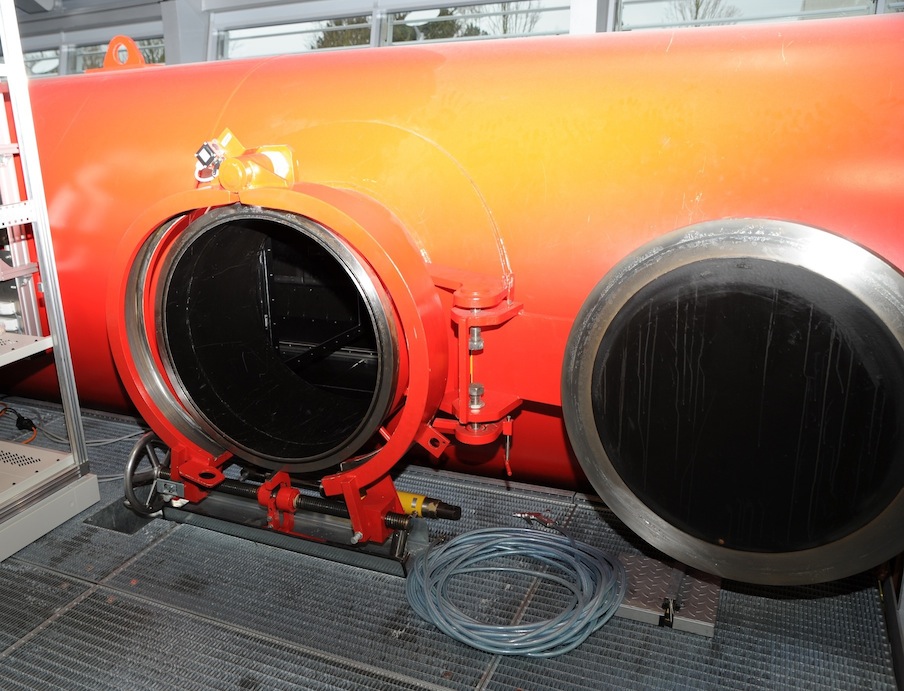}}}
\caption{Manhole at the end of one of the two measurement sections.  
Its diameter is 80\,cm.  }
\label{fig:manholes}
\end{figure}

\begin{figure}
\centerline{\scalebox{1.33}{\includegraphics{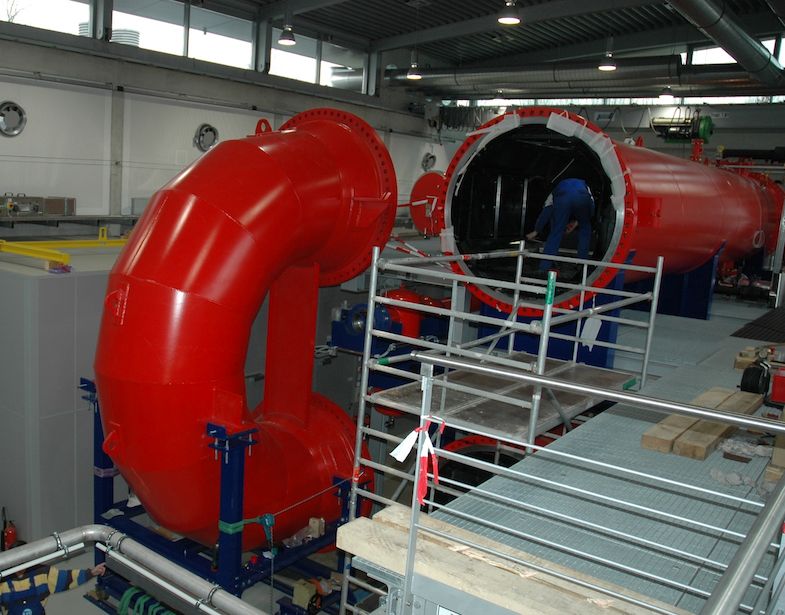}}}
\caption{One of two elbows is removed 
and reveals the cross section of the tunnel.  }
\label{fig:bend}
\end{figure}

\begin{figure}
\centerline{{\scalebox{0.3}{\includegraphics{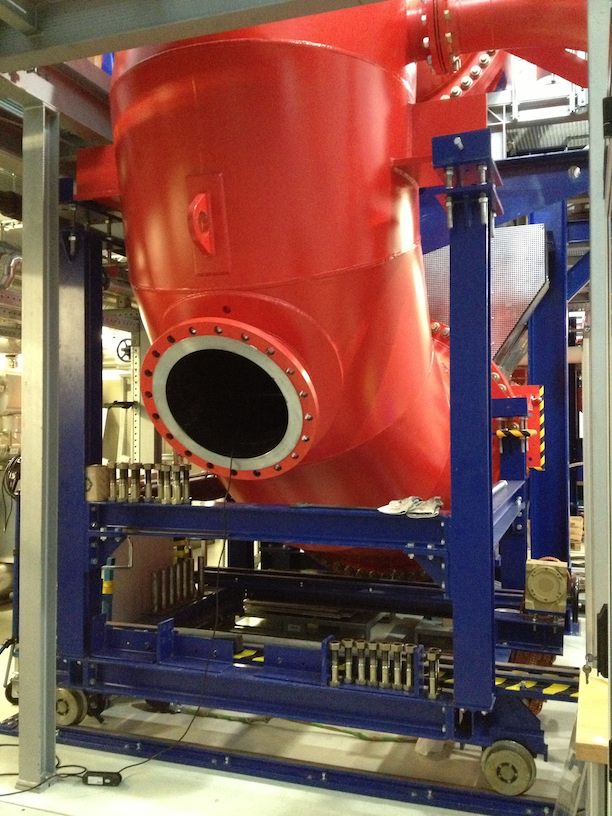}}}}
\caption{Additional manhole to access the space between the fan and the heat exchanger, 
and the frame on rails that supports the elbow when it is being moved.  }
\label{fig:frame}
\end{figure}

\begin{figure}
\centerline{\rotatebox{180}{\scalebox{0.3}{\includegraphics{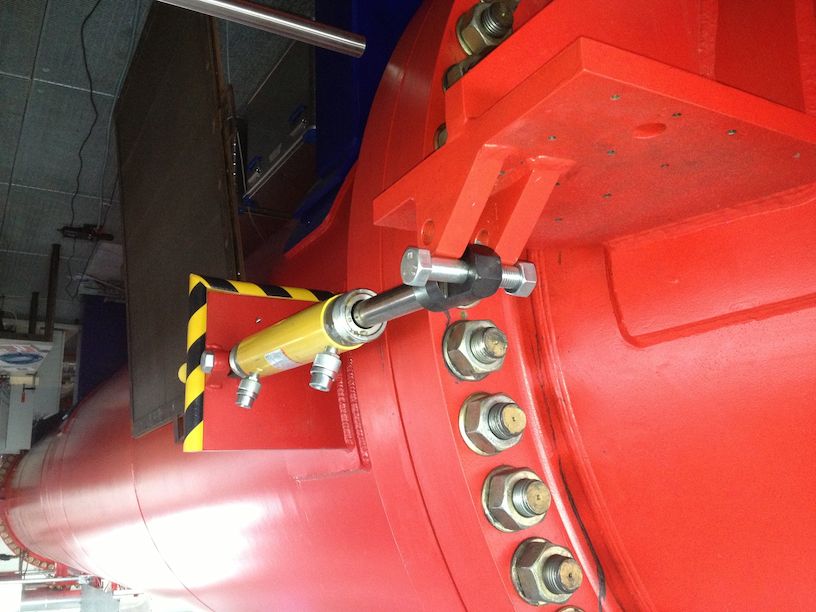}}}}
\caption{Hydraulic cylinders control the docking and undocking maneuvers of the elbows.  }
\label{fig:hydraulics}
\end{figure}

Mechanical access to the interior of the pressure vessel 
is possible both by removing the elbows, 
and through three manholes.  
Two of the manholes are on the sides of the two straight 
measurement sections (Fig.~\ref{fig:manholes}).  
These manholes are $\unit[0.8]{m}$ in diameter
and have quick-lock mechanisms to make it easy to open and close the tunnel.  
The additional manhole is placed in one of the elbows 
(see  Fig.~\ref{fig:frame}) 
to allow access to the space between the fan in the lower part and the heat exchanger 
in the upper part of the wind tunnel.  
This manhole is $\unit[0.6]{m}$ in diameter.  
To access the full cross section of the straight sections, 
the two elbows (Fig.~\ref{fig:bend}) can be removed with a movable frame (Fig.~\ref{fig:frame}).  
To ensure precise docking and undocking of the flanges between the elbows and the straight sections, 
the last few centimeters of the movement are controlled by four hydraulic cylinders 
(Fig.~\ref{fig:hydraulics}).  

The main flanges connecting the two elbows of the tunnel to the straight sections 
and the manholes all use re-usable O-ring seals.  
The tunnel's main flanges use double O-rings in rectangular grooves to minimize leakage, 
the manholes at the measurement sections use single O-rings in rectangular grooves, 
and the manhole at the bend uses a single O-ring in a trapezoidal groove.  
All other seals are kammprofile gaskets.  
The maximum leak rate of the pressure vessel and all connected parts of the entire facility are specified 
to be better than $\unit[0.5]{\%}$ of their respective volumes per year.  

% new sketch to update cross sectional area: 

\subsection{Working Fluids}

The wind tunnel can be filled and operated with any non-corrosive gas.  
We have so far used air, nitrogen and SF$_6$, but other gases are possible.  
% changes in response to the referee: 
The viscosity of SF$_6$, given in Table~\ref{tab:sf6visc} for various pressures, 
%At room temperature and pressure, 
%SF$_6$ has a density of $\rho = \unit[6.52]{kg/m^3}$ and a kinematic viscosity of 
%$\nu = \unit[2.32 \times 10^{-6}]{m^2/s}$.  
%These values change to $\rho = \unit[112]{kg/m^3}$ and $\nu = \unit[0.143 \times 10^{-6}]{m^2/s}$ 
%at $\unit[15]{bar}$,\cite{Hoogland1985} 
%which 
%make it possible to access a wide range of Reynolds number by changing the pressure alone.  
makes it possible to reach high Reynolds numbers at moderate pressures.  
% end changes.  
SF$_6$ is an inert gas, but decomposes above $\unit[1200]{^{\circ}C}$,\cite{Wilkins1969} 
though this temperature can be lower on catalytic surfaces, which include certain metals.  

%new table for reviewer: 

\begin{table}
\caption{Kinematic viscosity and density of SF$_6$ at various pressures 
and at \unit[295]{K}.\cite{Hoogland1985}  
}
\label{tab:sf6visc}
\begin{tabular}{l|cccccccc}
$p$ [bar]                                    & 1 & 3 & 5 & 7 & 9 & 11 & 13 & 15 \\
$\rho$ [kg/m$^3$]                      & 6.0 & 18.5 & 31.7 & 45.6 & 60.5 & 76.5 & 93.7 & 112.7 \\
$\nu$ [m$^2$/s] $\times 10^7$  & 25.1 & 8.2 & 4.8 & 4.2 & 3.2 & 2.0 & 1.7 & 1.4 \\
\end{tabular}
\end{table}

%end changes due to reviewer comment 1)

\subsection{Gas Handling System}

\begin{figure}
%\centerline{\scalebox{0.24}{\includegraphics{pictures/DSC_0169.jpg}}}
\centerline{\scalebox{0.96}{\includegraphics{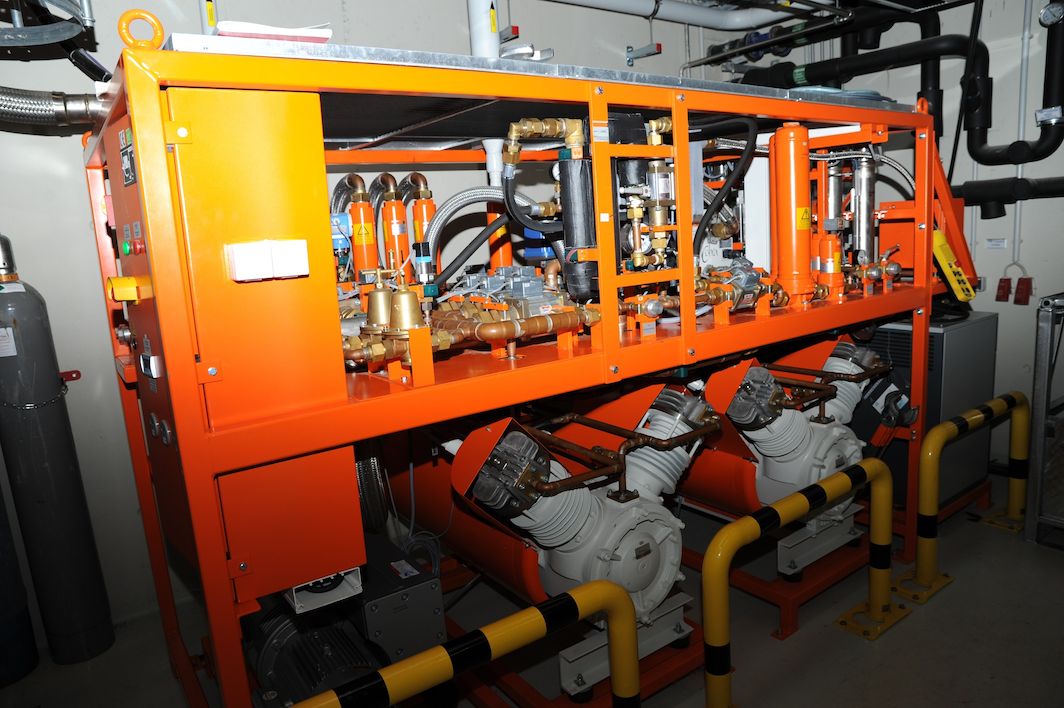}}}
\caption{Gas-handling system, including a vacuum pump and two compressors.  }
\label{fig:gashandlingsystem}
\end{figure}

\begin{figure}
%\centerline{\scalebox{0.24}{\includegraphics{pictures/DSC_0170.jpg}}}
\centerline{\scalebox{0.96}{\includegraphics{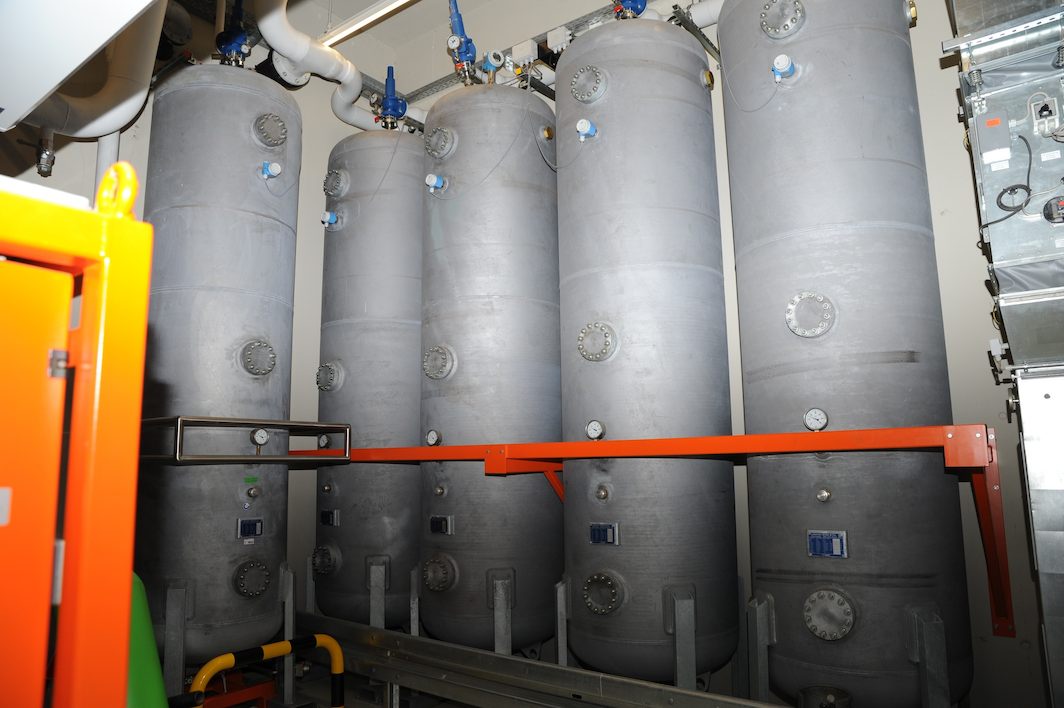}}}
\caption{Storage tanks, in which the SF$_6$ is stored in liquid form.  }
\label{fig:tanks}
\end{figure}

The facility includes an automated gas handling system (Fig.~\ref{fig:gashandlingsystem}).  
The system supplies the wind tunnel with SF$_6$ 
and other gases including dried air.  
The SF$_6$ is stored in the liquid phase in tanks (Fig.~\ref{fig:tanks}).  
The system both evacuates and pressurizes the wind tunnel.  
It also gasifies, liquefies, and cleans the SF$_6$.  

The typical cycle to prepare the wind tunnel for a run 
starts by closing the vessel to evacuate it.  
The evacuation is performed to minimize the amount of residual air 
in the tunnel.  
Once a pressure of $\unit[1]{mbar}$ has been reached, 
the tunnel is filled with the desired gas up to $\unit[15]{bar}$.  
The system can maintain a given pressure during measurements.  
When SF$_6$ is used, and after the run is complete, 
the system pumps the gas back to the storage tanks, 
and reduces the pressure in the vessel to $\unit[1]{mbar}$.  
The vessel is then filled again with air to $\unit[1]{bar}$ and opened.  

\subsection{Safety System}

All components of the facility have appropriate safety systems.  
The safety equipment includes over-pressure valves and burst plates, 
which prevent pressures from rising so high that they could damage parts of the system.  
To avoid contamination of the laboratory space with SF$_6$, 
all safety valves open into a pipe that ends outside the building.  
In addition, many components of the system have sensors that detect malfunctions 
and can shut down the entire system.  
For example, there is a sensor that shuts the motor down 
if the temperature inside the wind tunnel exceeds $\unit[40]{\tccelsius}$.  

\begin{figure}
%\centerline{\scalebox{0.24}{\includegraphics{pictures/DSC_0162.jpg}}}
\centerline{\scalebox{0.96}{\includegraphics{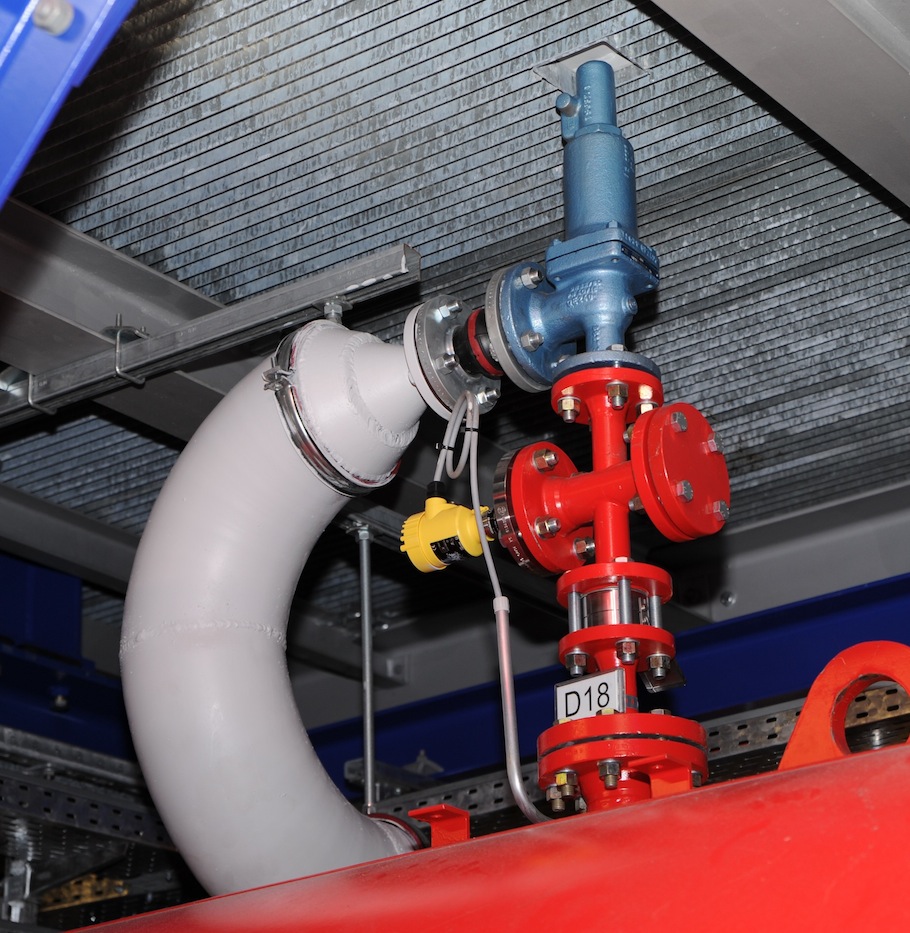}}}
\caption{Burst plate and safety valve with pressure monitor and exhaust pipe.  }
\label{fig:savetyvalve}
\end{figure}

The pressure vessel itself has a safety release system consisting of a burst plate and a safety valve 
(Fig.~\ref{fig:savetyvalve}) that limits the operational pressure to $\unit[15]{bar}$.  
At this pressure  the burst plate first breaks without releasing any SF$_6$ through the safety valve.  
Next, a gauge that monitors 
the pressure in the space between the burst plate and the safety value 
indicates an overpressure.  
Only if the pressure increases beyond 19\,bar does the safety valve release the gas 
from the pressure vessel.  
The wind tunnel itself is certified up to $\unit[20]{bar}$, 
and the operational pressure can be increased up to that value, 
provided the safety release system is redesigned in such a way that evacuation 
of the gas is guaranteed 
even when there is energy being injected into the system by, for example, the motor.  
There is a similar safety release system on the filter bypass (described in the next section), 
which opens at 19.5\,bar.  
Each of the two safety valves are connected with pipes that open outside the building.  
Flexible couplers are installed between the valves and the pipes 
in order to allow for vibration and misalignment.  

\begin{figure}
%\centerline{\scalebox{0.24}{\includegraphics{pictures/DSC_0168.jpg}}}
\centerline{\scalebox{0.96}{\includegraphics{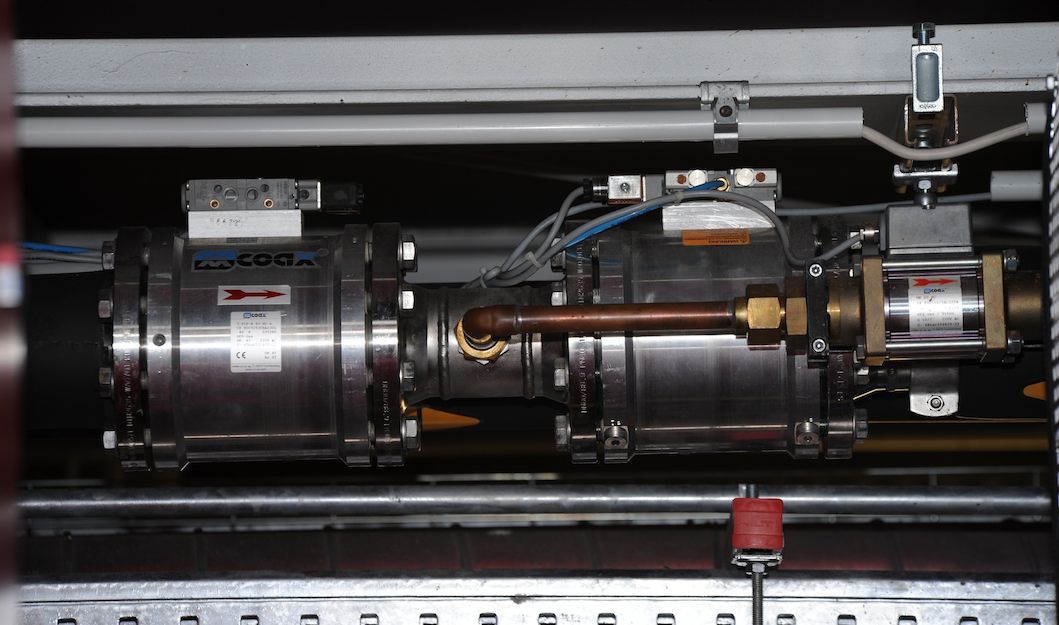}}}
\caption{Double-block-and-bleed valves in the gas supply line.  }
\label{fig:blockandbleed}
\end{figure}

During maintenance or installation of experimental equipment 
the safety of persons working in the wind tunnel must be guaranteed.  
Therefore, after an experiment and after the pressure vessel is again filled with air, 
the manholes are opened 
and the gas handling system is decoupled from the pressure vessel 
by double block-and-bleed valves (Fig.~\ref{fig:blockandbleed}).  
These are two valves in series blocking the gas supply pipe from the pressure vessel, 
and one valve opening the space between the two valves 
to release to the environment any gas that leaks.  
Furthermore, the wind tunnel is actively ventilated while open, 
and persons working in the wind tunnel carry O$_2$ detectors.  
Finally, portable SF$_6$ detectors are used to locate leaks 
and to check the SF$_6$ concentration in the pressure vessel before entering it.  

The experimental hall itself and the SF$_6$ storage room are both equipped with fire, 
SF$_6$ and O$_2$ detectors linked to a centralized alarm 
that initiates an evacuation in all dangerous situations.  
These alarms are also linked to the fire department 
in order to get fast and professional help in the case of dangerous situations.  

\subsection{Filter Bypass}

\begin{figure}
%\centerline{\scalebox{0.24}{\includegraphics{pictures/DSC_0150.jpg}}}
\centerline{\scalebox{0.60}{\includegraphics{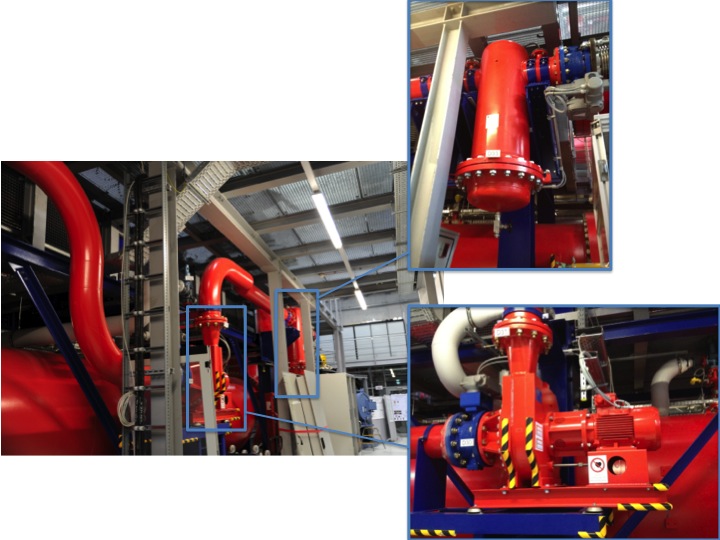}}}
\caption{Filter by-pass line with magnetically coupled pump (middle)
and F9 filter cassette (end).  
}
\label{fig:bypass}
\end{figure}

\begin{figure}
%\centerline{\scalebox{0.24}{\includegraphics{pictures/DSC_0151.jpg}}}
\centerline{\scalebox{0.96}{\includegraphics{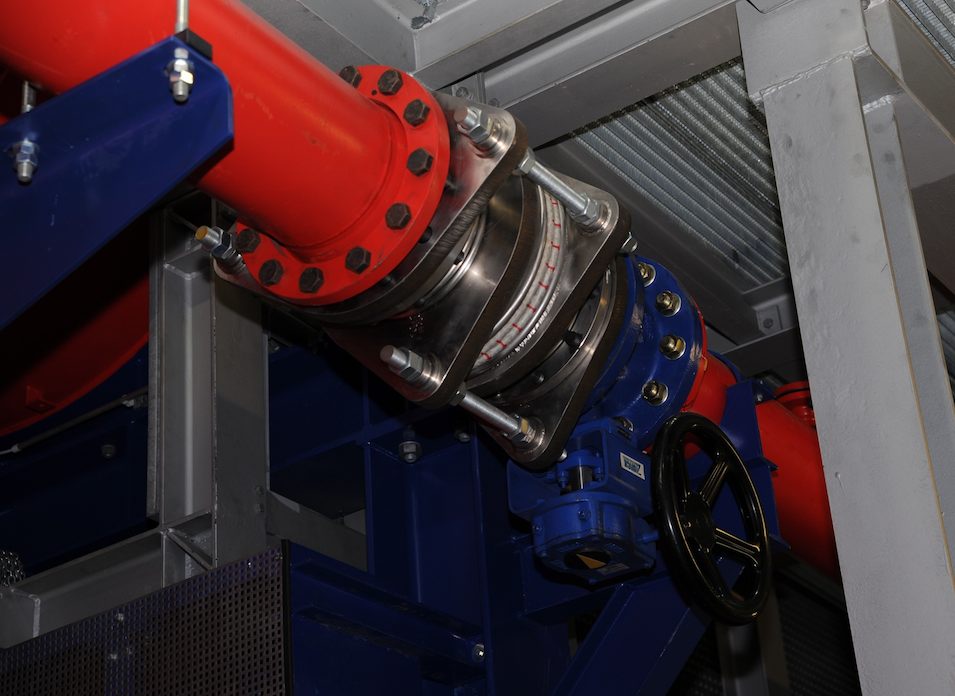}}}
\caption{One of two flexible couplers along the filter by-pass line.  }
\label{fig:bypass_coupler}
\end{figure}

The wind tunnel is equipped with a bypass to clean the gas in the pressure vessel.  
A pump draws gas from one elbow (see Fig.~\ref{fig:tunnel_sketch}), 
pushes it through a class F9 filter to remove particles from the gas, 
and returns it to the other elbow.  
The F9 filter efficiency is 98\% for $\unit[1]{micron}$ particles.  
The bypass itself is a $\unit[19]{m}$ long tube (Fig.~\ref{fig:bypass}) 
with an inner diameter of $\unit[250]{mm}$.  
The flow rate through the bypass is up to $\unit[400]{m^3/h}$.  
To avoid leaks the pump is entirely encapsulated, 
its connection to an external motor made through a magnetic coupling.  
To compensate for mechanical stresses caused by differences in the expansions of the bypass 
and the main body of the pressure vessel, 
the filter bypass includes two flexible couplers (Fig.~\ref{fig:bypass_coupler}).  
The expansion is due to temperature or pressure fluctuations.  
To prevent pressure-driven flow through the filter bypass during measurements, 
the bypass can be closed with any of three valves along its length. 
%comment 3
During the acquisition of the data presented here, 
the bypass was closed.  
The bypass was opened only to clean the gas before running the experiments.  
% end comment 3
Since the gas in the bypass can be enclosed by any pair of valves, 
an additional safety release consisting of a burst plate 
and a safety valve with pressure monitoring was necessary here.  

\subsection{Fan}
\label{sec:fan}
The pump that recirculates the gas in the wind tunnel is a $\unit[210]{kW}$ electric motor 
coupled to a fan with 20 blades.  
There is a stator with 17 blades, 
and the annular passage through both fan and stator 
has an inner diameter of $\unit[0.935]{m}$ 
and an outer diameter of $\unit[1.5]{m}$.  
Both fan and motor are inside the pressure vessel, 
in the downstream end of its lower section.  
The flow is ducted into the fan by guide plates upstream of the fan.  
The fan speed is controlled by a frequency controller, 
yielding a constant mean flow velocity that is adjustable 
between $\unit[0.5]{m/s}$ and $\unit[5]{m/s}$ with SF$_6$ at $\unit[15]{bar}$.  

The motor is housed to prevent particles added to the flow from causing damage to the bearings.  
The motor is water cooled, with  $\unit[40]{kW}$ cooling power available 
from the main cooling system of the facility.  
The cooling water for the motor is driven by a dedicated pump 
to ensure the minimum required flow rate necessary to cool the motor.  
If there were a leak in the cooling system, SF$_6$ could leak into it.  
Should this happen, the pressure increase in the system would be detected 
and valves in the cooling lines would close automatically.  

Although it is housed, the motor is exposed to the full range 
of pressures in the tunnel between vacuum and $\unit[20]{bar}$.  
For this reason, the motor coils were insulated by a specially made bubble-free coating.  
If bubbles were present, they could expand or contract under variations in pressure 
and damage the insulation.  

The motor temperature and vibration are monitored.  
The motor controller shuts the motor down when any problems are detected.  
All connections to the motor are made through leak-tight feedthroughs.  

\begin{figure}
%\centerline{\scalebox{0.0815}{\includegraphics{pictures/DSC_3858.jpg}}}
\centerline{\scalebox{1}{\includegraphics{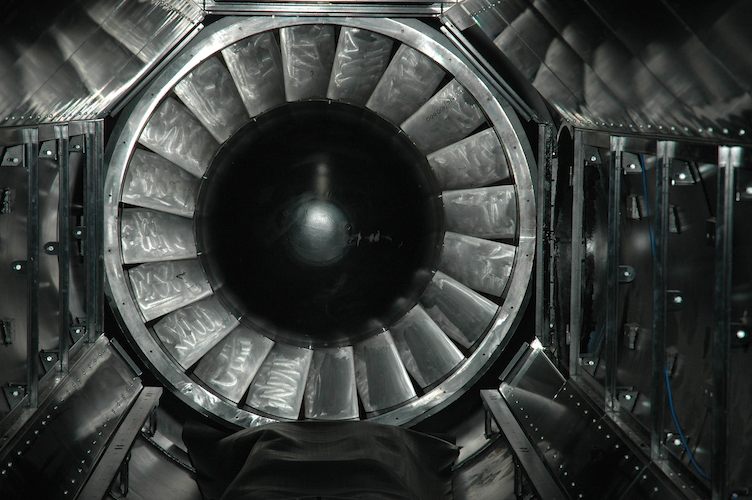}}}
\centerline{\scalebox{0.706}{\includegraphics{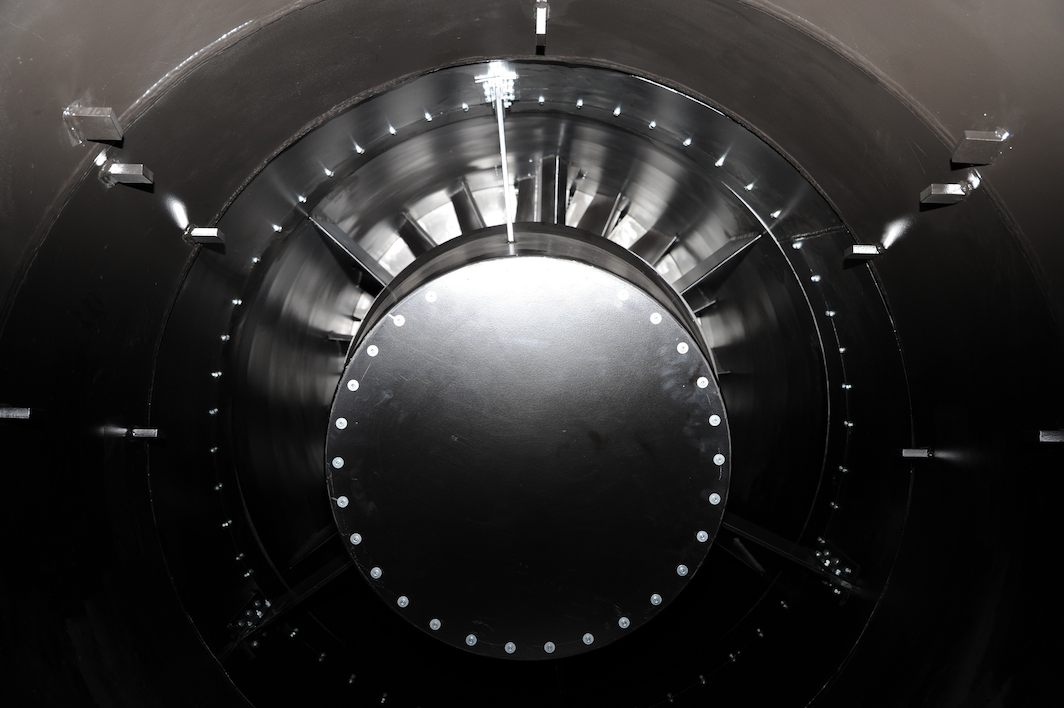}}}
\caption{Fan (looking downstream) and encapsulated motor (looking upstream) 
in the lower measurement section.  }
\label{fig:motorfan}
\end{figure}

\subsection{Heat Exchanger}
\label{sec:heatx}

\begin{figure}
%\centerline{\scalebox{0.24}{\includegraphics{pictures/DSC_0184.jpg}}}
\centerline{\scalebox{0.96}{\includegraphics{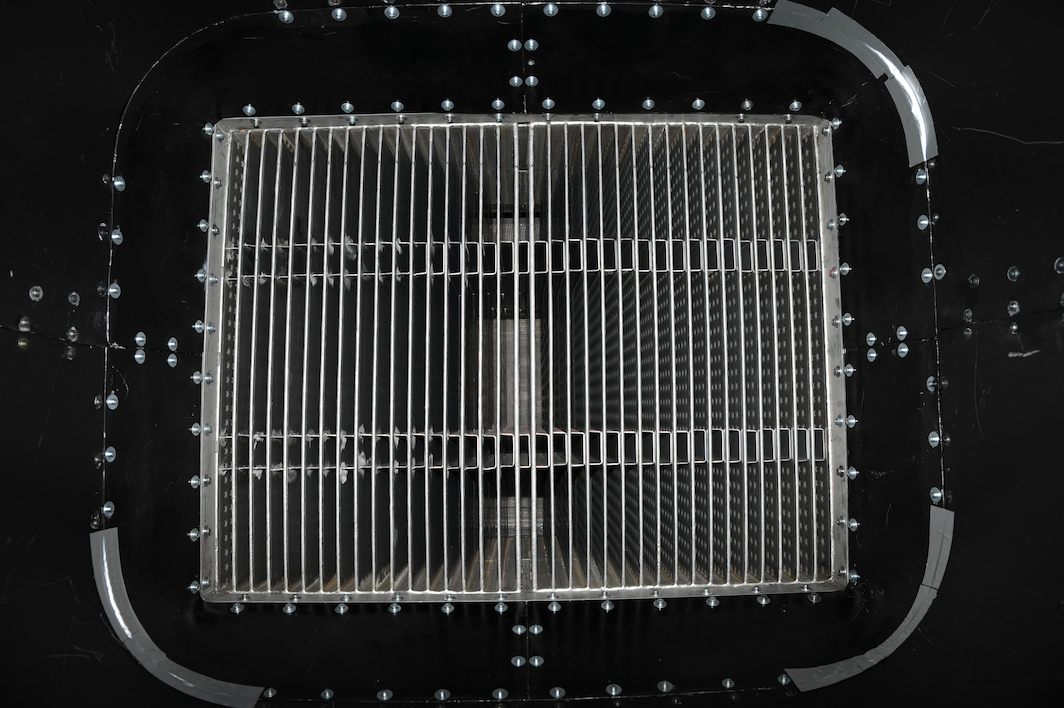}}}
\caption{Heat exchanger (looking downstream).  
Water flows through the vertical plates that we see here end-on, 
while the black shrouds guide the $SF_6$ through the heat exchanger.  }
\label{fig:heatexchanger}
\end{figure}

A heat exchanger inline with the flow 
removes the heat generated by pumping, 
\textit{i.e.} all of the mechanical energy put in by the fan.  
The heat exchanger consists of two water-cooled registers stacked in series.  
Each of these registers is $\unit[245]{mm}$ long and $\unit[1.268]{m}$ wide 
and consists of 36 vertical plates, 
as can be seen in Fig.~\ref{fig:heatexchanger}.  
The inside height is $\unit[0.96]{m}$.  
The plates are $\unit[5]{mm}$ thick with an open space of $\unit[28]{mm}$ between each plate.  
An earlier version of the heat exchanger suffered from aeroelastic resonances, 
which were suppressed by the addition of U-shaped spacers in the open space between the plates.  
The heat exchanger is centered with respect to the wind tunnel.  
The flow is guided from the elbow into the heat exchanger by shrouds, 
so that there is no flow around the heat exchanger.  

The incoming cooling water flows through one half of the heat exchanger (18 plates), 
gets collected, turns back and flows through the other half of the heat exchanger.  
To avoid temperature gradients over the tunnel's cross-section, 
the cooling water flows through the two heat exchangers in the same rotational orientation 
but the two heat-exchangers are rotated by $\unit[180]{\tcdegree}$ with respect to each other.  

To achieve maximum efficiency, 
the heat exchanger is connected directly to the cooling water system of the building; 
there is no additional heat exchanger.  
The two registers of the heat exchanger receive the same constant flow rate, 
which is driven by a common pump.  
The temperature controller holds 
constant the fluid temperature to set points between $\unit[20]{\tccelsius}$ and $\unit[35]{\tccelsius}$.  
This is realized by mixing cold water from the building with 
hot water from the heat exchanger return flow.  
Since this causes a varying flow rate in the support line from the building, 
an additional pump on the building side in combination with an excess flow valve 
realizes a constant flow rate on the building side.  

Since the cooling water is at the nominal pressure of the building water supply, 
there is a pressure difference across the surfaces of the heat exchanger.  
The heat exchanger is designed 
to sustain the full pressure difference between the cooling water 
and the pressure inside the wind tunnel (between vacuum and maximum $\unit[20]{bar}$ absolute).  
The cooling water flows to the heat exchanger through feedthroughs 
that on one hand are flexible enough 
to accommodate deformations due to temperature and pressure variations, 
and on the other hand are stiff enough to withstand the full pressure difference between 
the pressure of the cooling water system inside and the wind tunnel pressure on the other side.  

\begin{figure}
%\centerline{\scalebox{0.24}{\includegraphics{pictures/DSC_0156.jpg}}}
%\centerline{\scalebox{0.08}{\includegraphics{pictures/IMG_0990.jpg}}}
\centerline{\scalebox{0.25}{\includegraphics{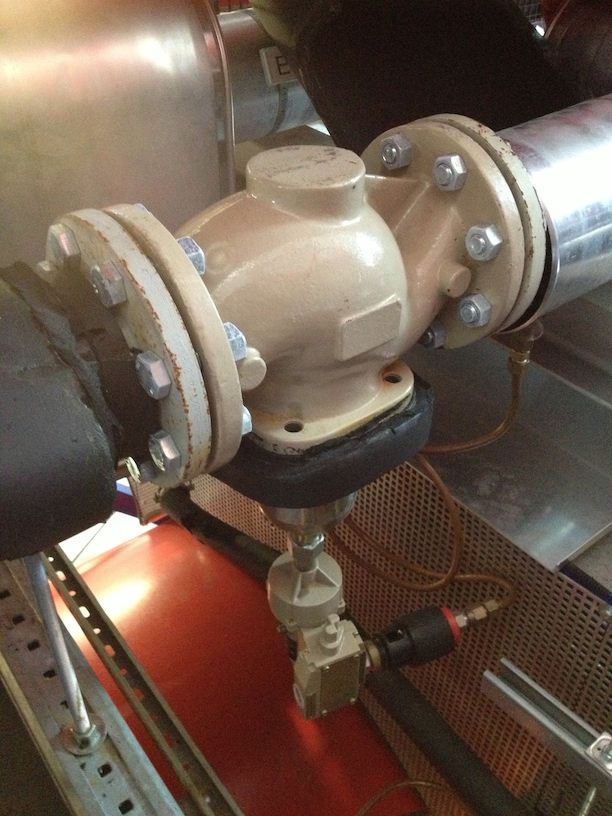}}}
\caption{Safety valves in the cooling water pipes.  }
\label{fig:coolingvalves}
\end{figure}

To prevent SF$_6$ from penetrating the cooling water system of the building 
in the case of a leak in the heat exchanger, 
the cooling water supply pipes include two valves that close at a pressure of $\unit[4]{bar}$ 
over the nominal.  
An additional four valves 
(see Fig.~\ref{fig:coolingvalves}) 
close at an over-pressure of $\unit[6]{bar}$.  
In the case of a leak, 
the pressure in the building system first increases, 
which causes the two valves in the supply pipes to close 
and isolate the building side of the cooling system.  
Then the pressure continues to increase until the four valves at the wind tunnel close 
in order to enclose the SF$_6$ in the wind tunnel.  
In the case of a leak of water into the tunnel, 
a conductive humidity sensor placed below the heat exchanger initiates an alarm 
and shuts down both the wind tunnel fan and the cooling system.  

\subsection{Measurement Sections}
\label{sec:exp}

\begin{figure}
%\centerline{\scalebox{0.25}{\includegraphics{figures/crosssection.png}}}
\centerline{\scalebox{0.33}{\includegraphics{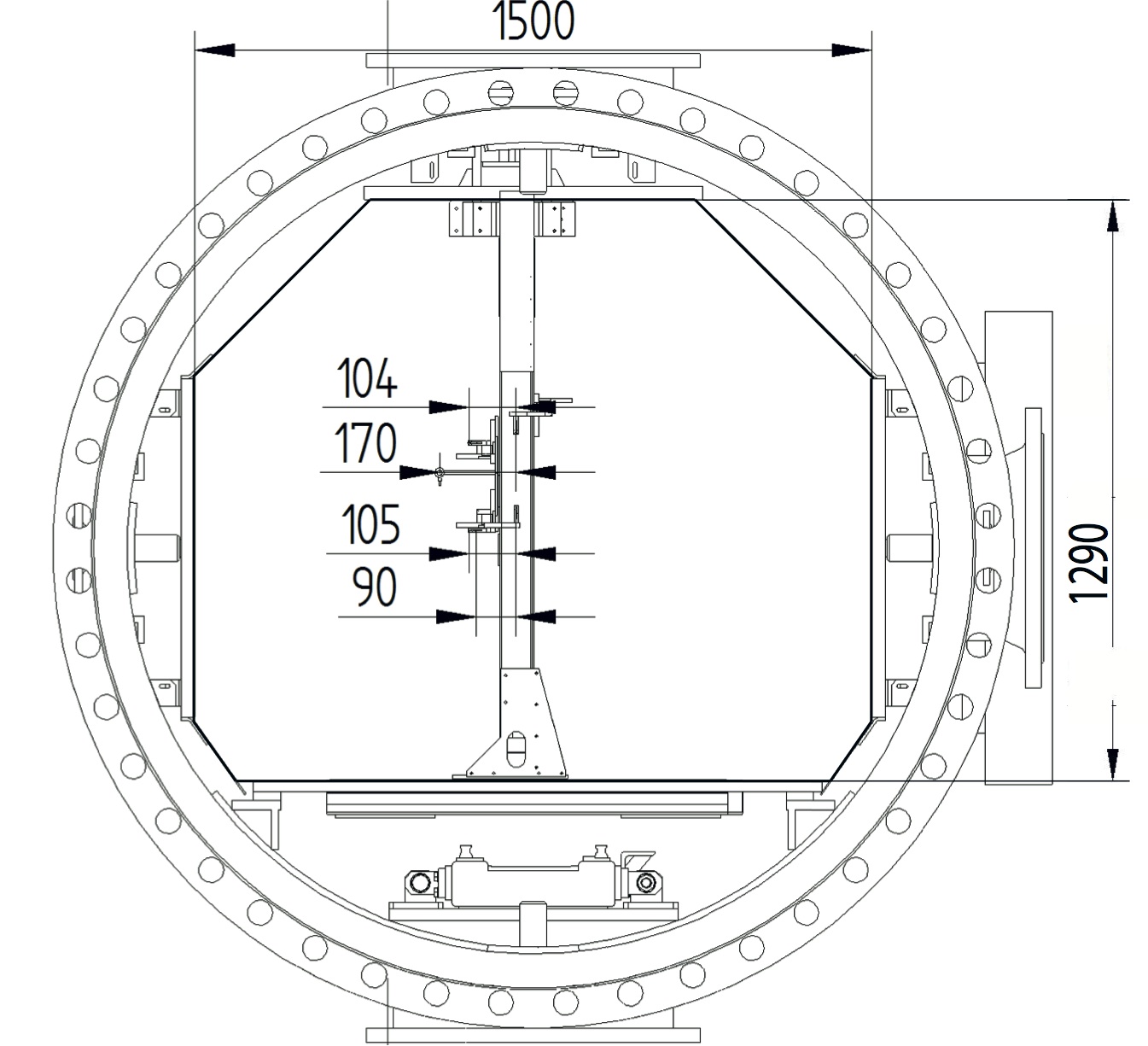}}}
\caption{
Cross section of the wind tunnel.  
Also shown is the linear traverse that held the hot wires, discussed below.  
Dimensions in \emph{mm}.  
}
\label{fig:crosssection}
\end{figure}

The long straight sections of the pressure vessel contain the measurement sections.  
The wind flows through interior ducts, or wind tunnels, as seen in cross section 
in Fig.~\ref{fig:crosssection}.  
The width of the tunnels are about 1.5\,m, their heights are about 1.3\,m, 
and the corners of their cross sections are cut off to make them approximately octagonal.  
The tunnels have a cross-sectional area of $\unit[1.7]{m^2}$, 
which slightly increases downstream through the inclination the roof by $\unit[0.114]{\tcdegree}$ 
and the side walls by $\unit[0.057]{\tcdegree}$ in the upper measurement section, 
and $\unit[0.110]{\tcdegree}$ and $\unit[0.055]{\tcdegree}$, respectively, 
in the lower measurement section.  
The upper measurement section is 8.8\,m long, and the lower one 6.8\,m long.  

At the upstream end of each test section, 
there are four mounts, one each on the top, bottom, left and right sides, 
to which turbulence generators can be mounted.  
Each mount can withstand forces up to 10\,kN continuously in the direction of the flow, 
and transients up to 22.5\,kN in the case of an accident.  
At the end of the $\unit[8.8]{m}$ long upper measurement section 
a shroud adapts the cross section of the measurement section to the circular profile of the elbow.  
The same is true at the entrance of the lower measurement section.  
All inner parts of the wind tunnel are painted black 
to minimize reflections in optical measurements.  

\subsection{Electrical and Optical Access}

\begin{figure}
%\centerline{\scalebox{0.24}{\includegraphics{pictures/DSC_0153.jpg}}}
\centerline{\scalebox{0.96}{\includegraphics{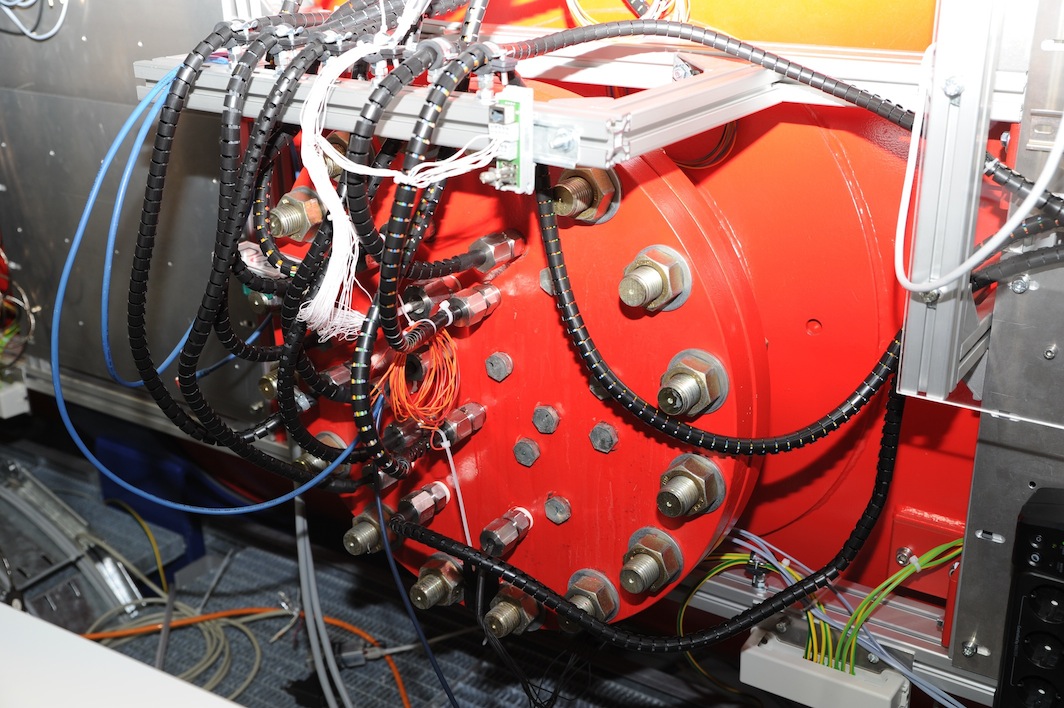}}}
\caption{Flange with multiple $\unit[3/4]{"}$ Conax NPT feed-throughs with up to 60 wires each.  }
\label{fig:feedthroughs}
\end{figure}

To connect electrical probes and other electrical equipment inside the wind tunnel to the outside, 
we use electrical feed-throughs.  
For this reason, each of the two measurement sections have two $\unit[400]{mm}$ flanges 
onto which plates with tapped holes are mounted.  
Each plate has 21 $\unit[3/4]{"}$ NPT taps.  
Depending on the required current through the wires, 
different kinds of feedthroughs with different numbers of wires 
and different wire diameters have been installed.  
Up to 60 signal wires can pass through each $\unit[3/4]{"}$ NPT tap (see Fig.~\ref{fig:feedthroughs}).  

\begin{figure}
\centerline{\scalebox{0.8}{\includegraphics{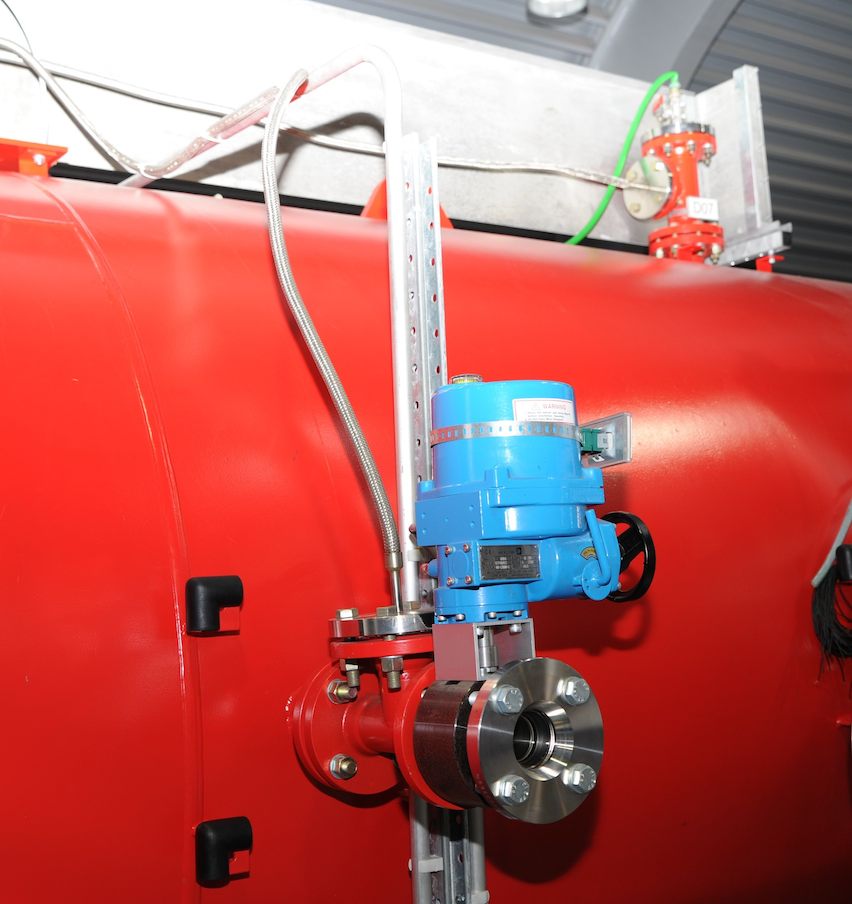}}}
%\centerline{\scalebox{0.24}{\includegraphics{pictures/DSC_0147.jpg}}}
\centerline{\scalebox{0.96}{\includegraphics{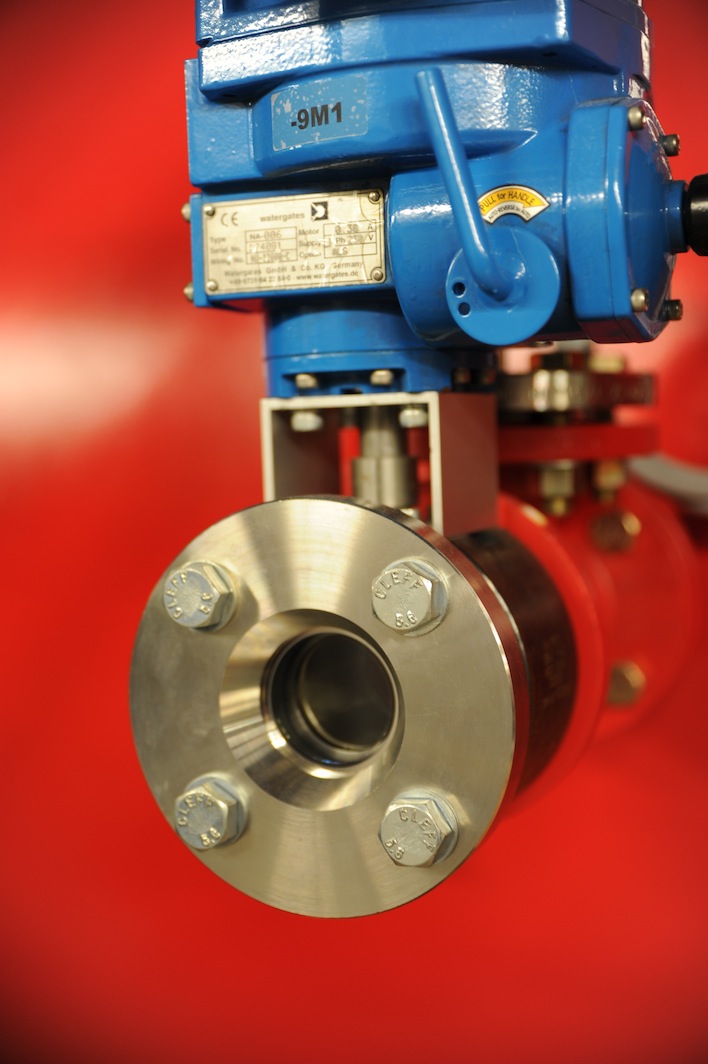}}}
%\centerline{\scalebox{0.24}{\includegraphics{pictures/DSC_0154.jpg}}}
\caption{One of two optical access windows with ball valve and dynamic pressure monitoring.  }
\label{fig:window}
\end{figure}

Optical access to high-pressure devices is generally difficult to implement.  
The VDTT has two borosilicate glass windows with an open diameter of $\unit[50]{mm}$ 
with a direct view into the wind tunnel (see Fig.~\ref{fig:window}).  
The outer diameter of the glass is $\unit[107]{mm}$, with thickness $\unit[20]{mm}$.  
To prevent the catastrophic release of SF$_6$ through a broken window, 
there are ball valves between the pressure vessel and the glass windows.  
These ball valves close automatically when there is a flow through the flange.  
The trigger for the ball valves to close is the detection of 
a pressure difference between the flange and the interior of the tunnel, 
which is zero unless there is a flow through the flange.  

Aside from these two windows, 
optical access also is possible through optical fibers, which pass through feedthroughs.  
A $\unit[150]{W}$ Nd:YAG laser has been coupled into such a fiber for illumination 
of the test section for optical particle tracking.  

\subsection{Aerodynamic Considerations}
\label{sec:aero}

Here we describe in general terms the character of the flow as it circulates through the tunnel.  
One of our main interests was to produce a homogeneous flow in the measurements sections.  
Residual inhomogeneity of the flow is set by the tunnel geometry upstream of the sections.  
The flow is sustained by the fan described in section \ref{sec:fan}, 
where it will acquire some rotation depending on the fan speed, 
since the stator has a fixed geometry.  
Thereafter, the flow passes through the elbow described in section \ref{sec:vessel}.  
The elbows contain no vanes to guide the flow.  
There are, however, mounting flanges in the elbows so that guide vanes can be 
installed if required.  
As described by G\"ortler \cite{goertler:1940} and Hawthorne \cite{hawthorne:1951}, 
oscillating secondary flows probably emerge in these elbows 
due to their curvature.  

The flow then enters the heat exchanger (Section \ref{sec:heatx}) 
through a contraction with an area ratio of 1.5 
that smoothly adapts the circular cross section of the elbow 
to the rectangular cross section of the heat exchanger.  
From the point of view of the flow, 
the heat exchanger consists of vertical slots.  
These slots destroy large-scale vortical structure.  
We observed, however, that the flow exited 
more quickly from the top of the heat exchanger 
than from the bottom of it.  

\begin{figure}
\centerline{\scalebox{0.48}{\includegraphics{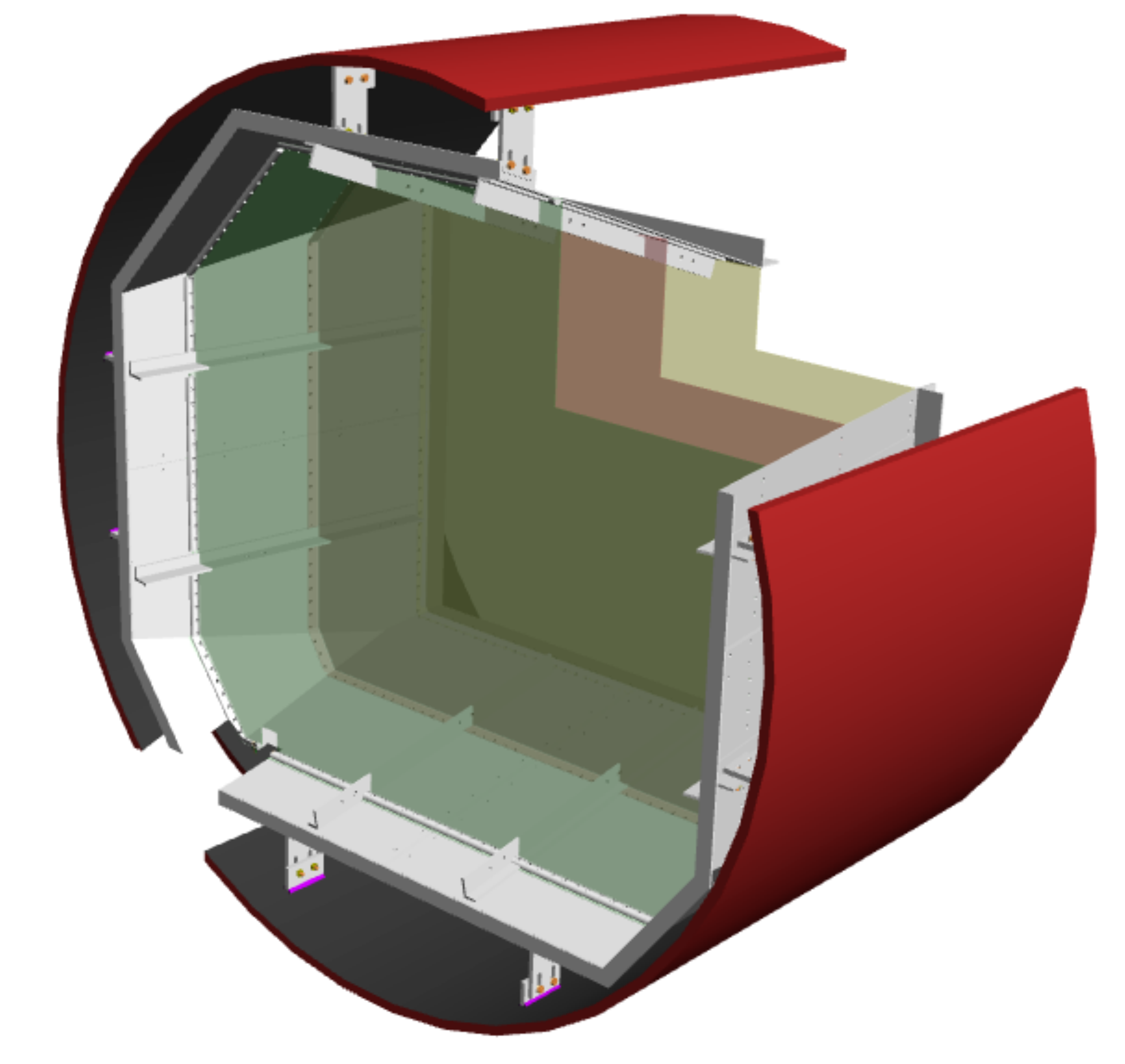}}}
\caption{Drawing of the expansion between the heat exchanger and the upper measurement section.  
The flow emerges from the face of the expansion in this view, so that the flow goes from right to left.  
The screens are shown as colored planes, with a cut-out for improved visibility.  }
\label{fig:expansion}
\end{figure}

Between the heat exchanger and the entrance to the upper measurement section 
is an 800\,mm long expansion that adapts the cross section of the heat exchanger, 
which has an area of 1.2\,m$^2$, 
to the cross section of the measurement section, 
which has an area of 1.7\,m$^2$ 
(Fig.~\ref{fig:crosssection}).  
As can be seen in Fig.~\ref{fig:expansion}, 
the walls of the expansion consist of flat plates, 
so that separation can occur at the corners along the inlet edge of the expansion.  
We observed this separation with telltales and cameras.  
In order to stabilize and homogenize the flow in the expansion,\cite{schubauer:1948} 
we added screens at approximately 250\,mm intervals 
along the length of the expansion.  
The meshes had the following properties, in the order encountered by the flow: 
wire diameter in $\unit{mm}$/wire mesh spacing in $\unit{mm}$: 
0.4/0.850, 0.4/1.267, 0.5/2.833.  

As described below in Sec.~\ref{sec:grids}, 
grids at the upstream end of the measurement sections produced turbulence.  
The grids were 545\,mm downstream of the last screen in the expansion described above.  
The absence of large scale rotation in the flow in the test section 
was checked with a swirl meter.  
A more detailed view of the quality of the flow in the test sections can be found 
in Section \ref{sec:testexp}.  

After the upper measurement section, the flow turns in the elbow, 
and returns to the fan through the lower measurement section.  
Between the elbow and the lower measurement section are 
a section that adapts the cross section of the elbow to that of the measurement section, 
and a bank of wire-mesh screens with properties similar to the ones 
in the expansion upstream of the upper measurement section.  
Note that there is no heat exchanger, nor expansion upstream 
of the lower measurement section.  

\subsection{Grids}
\label{sec:grids}

We introduce here an active grid that we designed for the VDTT, and which is under further development.  
We also describe the classical grids that we employed 
to produce the turbulence we characterize in the next section.  

Active grids were developed as a way to generate in wind tunnels 
high-Reynolds-number flows 
with convenient properties.\cite{makita:1991}  
Active grids work by stirring the flow with rotating paddles, 
rather than disturbing it through the wakes of stationary bars, as in a classical grid.  
Modern active grids generate not only high-Reynolds-number flows, 
but also flows with tailored properties.\cite{cekli:2010}  
Such control is desirable where turbulence with certain statistical properties is needed, 
as is the case when the atmospheric boundary layer needs to be synthesized 
to observe its effect on wind turbines,\cite{knebel:2011} 
or where the effects of variation of the large-scale properties of the turbulence 
on the small-scale dynamics need to be understood.\cite{blum:2011}  
Since the large-scales are created by the geometry of the apparatus in an experiment, 
one advantage of the active grid is that its geometry is variable and can be adjusted during its operation.  
In this way, the response of the turbulence 
to changes in the properties at the large scales can be measured.  

\begin{figure}
%\centerline{ \hbox{ \vbox{ \hbox{ \scalebox{0.37} {\includegraphics{pictures/paddle.jpg}} } \vbox to 7mm {\hbox{}} }} \hfil \hbox{ \vbox{ \hbox{ \scalebox{0.3} {\includegraphics{pictures/active_grid.jpg}} } } } }
\centerline{\scalebox{0.33}{\includegraphics{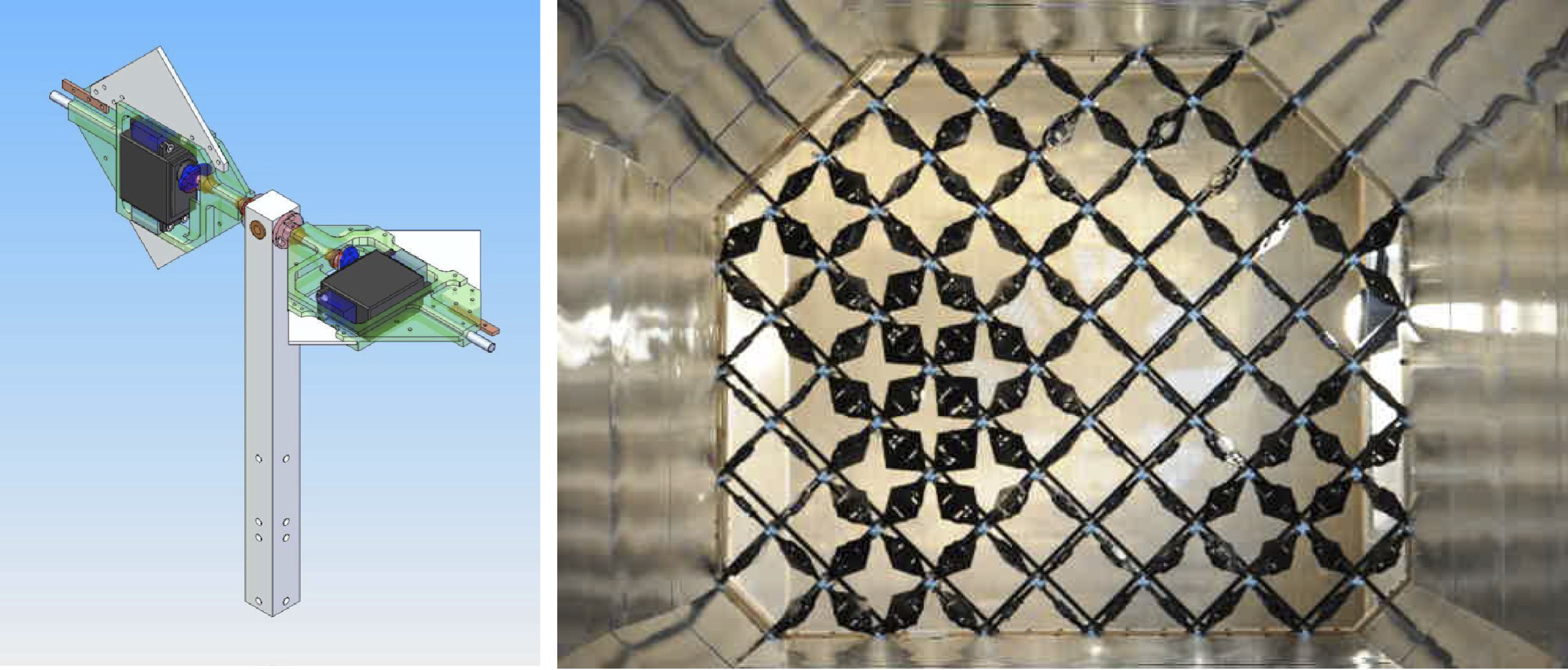}}}
\caption{On the left is a computer rendering of two active-grid paddles, 
each with its own integrated servo motor.  
129 such paddles compose the active grid, seen face-on in the right panel.  
Here, the paddles are at various angles to the flow, 
in some places blocking the flow more than in others.  
The grid is shown here installed in the Prandtl tunnel, 
which is itself shown in Fig.~\ref{fig:prandtltunnel} below.  }
\label{fig:activegrid}
\end{figure}

Our active grid advances the state of the art because there are many more degrees of freedom 
in the motions of its paddles than in previous grids.  
There are 129 degrees of freedom, whereas others had about 20.  
This gives an unprecedented level of control over the turbulence generated by the grid.  
Each degree of freedom corresponds to a single diamond-shaped paddle, 
the collection of which tile the cross section of the tunnel, as seen in Fig.~\ref{fig:activegrid}.  
Each paddle has its own computer-controlled servomotor 
that adjusts the angle of the paddle relative to the mean flow 
about an axis perpendicular to the flow.  
The paddles block the flow locally to a degree that depends on the angle of the paddle.  
The paddle angles can change over time according to some algorithm programmed by the user.  

\begin{figure}
\centerline{\scalebox{0.5}{\includegraphics{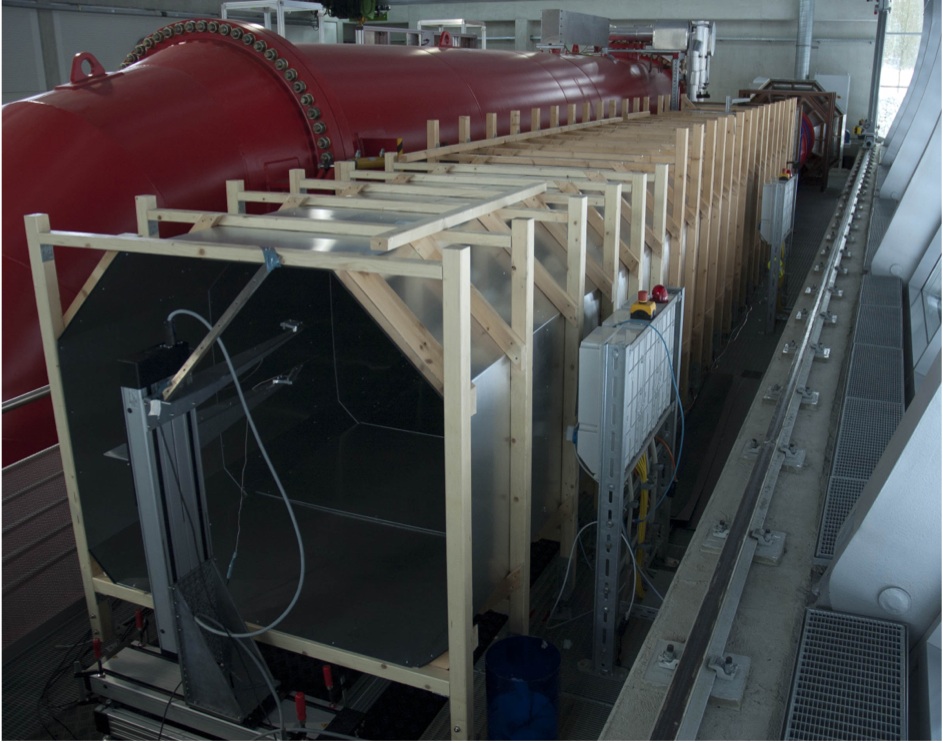}}}
\caption{The Prandtl tunnel sits parallel to the VDTT with a cross section identical 
to the one inside the VDTT.  
Most easily visible here, at the left side of the picture, is the outlet of the tunnel, 
with a linear traverse that holds hot wire probes.  
The wooden frames and sheet metal walls were built in 2011.  
The inlet and fan in the background, at the top right of the picture, 
are the parts built nearly 80 years ago.  
}
\label{fig:prandtltunnel}
\end{figure}

The active grid is being tested in an open-return wind tunnel \cite{kastrinakis:1983} built originally 
in the Kaiser-Wilhelm Institute of Ludwig Prandtl by Fritz Schulz-Grunow 
and Hans Reichhardt from 1936-1938.  
As seen in Fig.~\ref{fig:prandtltunnel}, 
the test section is 10\,m long, with a cross section identical to the one in the VDTT.  
The maximum flow speed in the Prandtl tunnel is approximately 11\,m/s.  
The outcome fulfill the requirements of increasing the turbulence from $R_\lambda \approx 300$ to $1500$. Details from  these tests will be presented in a separate paper.  

\begin{figure}
\centerline{\scalebox{1}{\includegraphics{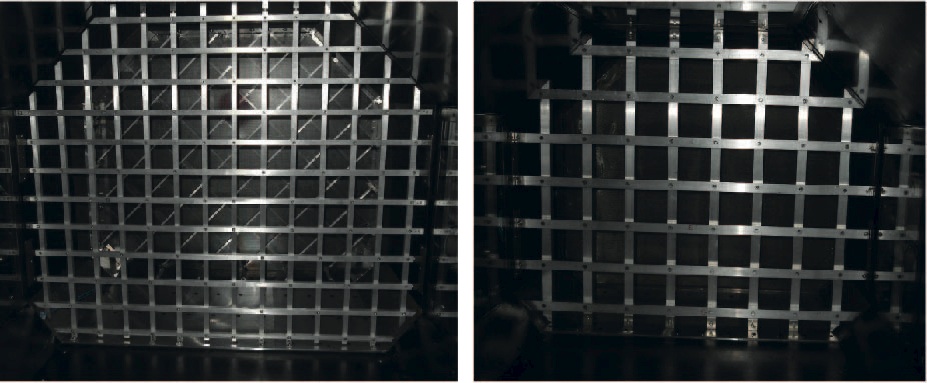}}}
\caption{Two of the classical grids we used to excite turbulence at the upstream end 
of the upper measurement section.  
Here, we look upstream at the 106.6\,mm grid on the left, 
and the 186.6\,mm on the right.  
Behind the 106.6\,mm grid is another diagonal grid and screen that we installed to control the flow, 
and which we subsequently removed.  
}
\label{fig:passivegrid}
\end{figure}

To make the measurements presented in this paper, 
we installed classical grid turbulence generators 
at the upstream end of the upper measurement section in the VDTT.  
We used one of three grids of traditional construction.\cite{comte-bellot:1966}  
They were composed of crossed bars with square cross sections in two layers, 
with the vertical bars being upstream of the downstream ones (Fig.~\ref{fig:passivegrid}).  
The distances between the bars of each grid were 53.3\,mm, 106.6\,mm and 186.6\,mm.  
The grid bars had widths 10\,mm, 20\,mm and 40\,mm, respectively, 
so that the grids blocked about 34\%, 34\% and 38\% of the area of the cross section.  
The grids were designed to have a distance of about a half mesh spacing 
between the wall and the first bar on each of the bottom, top, left and right sides.  
As in Bewley,\cite{bewley:2006} 
we found that the flow profile was sensitive to the precise geometry of the intersection 
between the grid and the wall.  
Because of this, we adjusted the width of the bars closest to the top and bottom walls 
in order to optimize the homogeneity of the flow in the center of the tunnel.  

\subsection{Traverses}

Within the VDTT, measuring instruments such as cameras 
will move down the length of the tunnel at the mean speed of the flow.  
This way, 
it is possible to follow the motions of particles within the flow, 
rather than sampling them as they sweep past a fixed position.  
That is, 
one can view turbulence from the Lagrangian, rather than Eulerian, perspective.  
The basic idea is that cameras take movies of particles that are suspended in the flow.  
A laser light source illuminates the particles.  
From views acquired by multiple cameras, 
software determines the three-dimensional positions of the particles at a series of instants, 
and then reconstructs their tracks through three-space over time.\cite{ouellette:2006}  
To accomplish this, the measurement sections 
accommodate movable sleds 
weighing up to 350\,kg that are driven by linear motors.  
The sleds will carry cameras, optics, and other instruments 
at the mean velocity of the circulating gas, up to 5\,m/s.  
There are few comparable installations,\cite{snyder:1971,sato:1987,ayyalasomayajula:2006} 
and none on the scale of the VDTT.  
The alignment of the cameras, 
the focusing of the lenses and the calibration of the imaging system 
will be performed automatically within the pressure vessel.  

A prototype of the linear motor for the wind tunnel has been built 
and is presently being tested outside of the tunnel.  
The platform that carries the high-speed cameras 
is driven back and forth along high-precision, low-friction rails by an electrical linear motor.  
The 88\,kW linear motor delivers up to 7.6\,kN of force to accelerate a payload 
(cameras and optics) 
at one end of the tunnel, carry them along the length of the tunnel, 
and decelerate them at the other end.  %\cite{zellman}.  
A linear encoder that has an accuracy of 0.2\,$\mu$m 
controls the position and the velocity of the platform.  

As in most wind tunnels, 
measurement equipment can otherwise be mounted anywhere along the measurement sections.  
Devices can be mounted directly to the walls, or on linear traverses.  

\begin{figure}
\centerline{\hbox{\vbox{\hbox{\scalebox{0.6}{\includegraphics{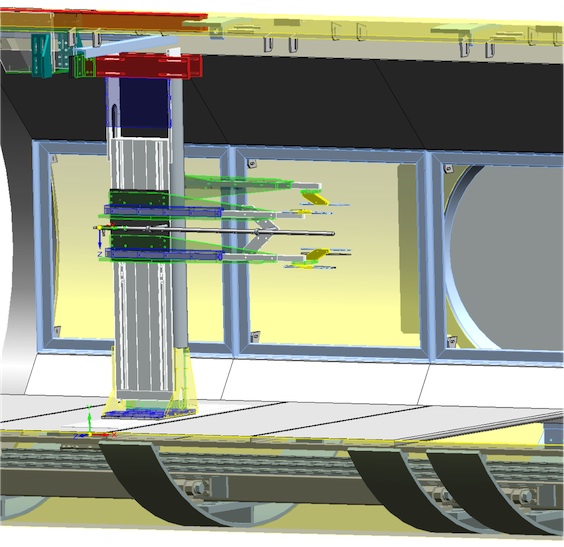}}}\vbox to 7mm{\hbox{}}}}\hfil \hbox{\vbox{\hbox{\scalebox{0.3}{\includegraphics{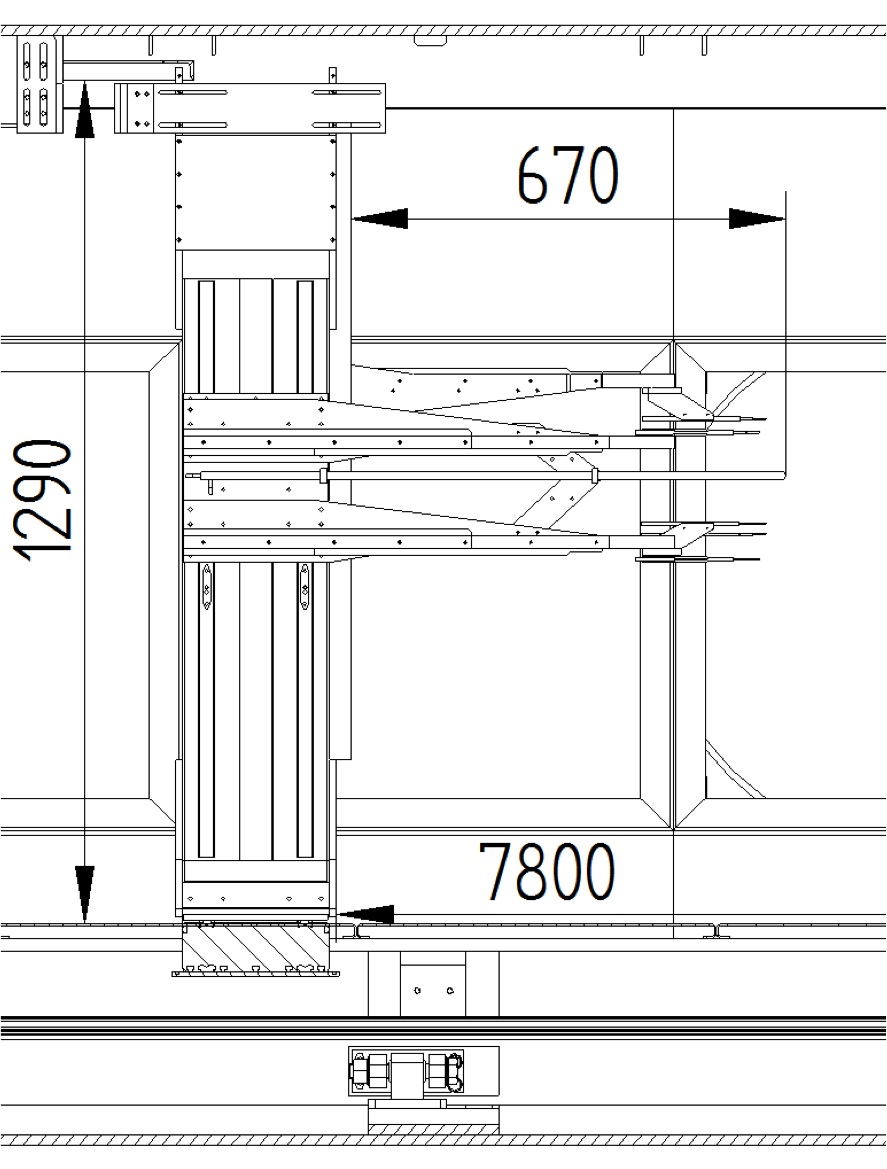}}}\vbox to 7mm{\hbox{}}}}\hfil \scalebox{0.8}{\includegraphics{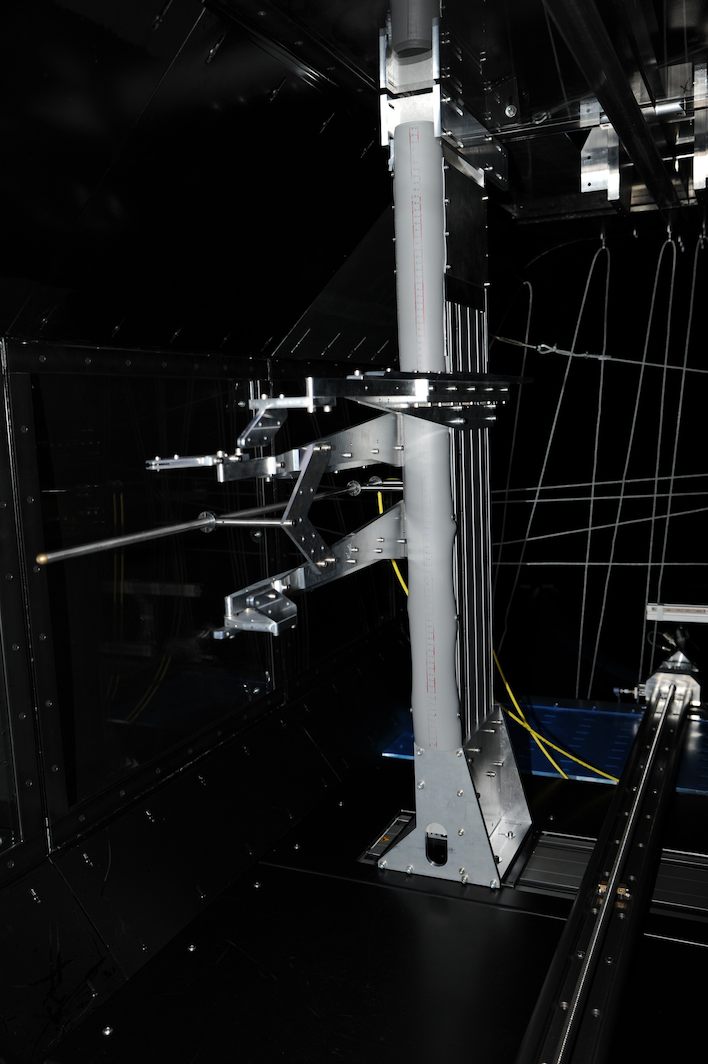}}}
\caption{A linear traverse for hot-wires, temperature sensors and Pitot-static tubes.  
Dimensions in \emph{mm}.  
The distance of 7800\,mm in the center panel is measured to the face of the grid.  
See also Fig.~\ref{fig:crosssection}.  }
\label{fig:traverse}
\end{figure}

The probes for the test measurements presented below 
were mounted on a 2D traverse, manufactured by Isel, 
to make possible measurements at different fixed positions 
in the cross-section of the tunnel, 
and at a single distance from the grid (Fig.~\ref{fig:traverse}).  
The traverse moved $\unit[0.65]{m}$ in the horizontal direction 
and $\unit[0.60]{m}$ in the vertical.  
The tips of the Pitot tube and the hot-wire probes were 
about $\unit[0.77]{m}$ upstream of the face of the traverse.  
The parts supporting the probes were designed according to the rule of thumb 
that the width of the part be about ten times smaller than distance of the 
part from the probe.  
This was to minimize the flow distortion caused by the probe supports, 
while still providing the stability needed to minimize probe vibration.  
For some of the measurements, one probe was held at a fixed height, 
so that the distance between it and the probes moving up and down on the vertical traverse 
could be varied.  
We did this to measure correlations between velocities separated across the width of the tunnel.

\subsection{Measurement Systems}

The arsenal of diagnostic equipment familiar in fluid mechanics 
can be used in the VDTT.  
We have designs to incorporate 
hot wire anemometry, 
cold wire thermometry, 
laser Doppler velocimetry, 
acoustic velocimetry, 
particle sizing, 
dynamic pressure measurement, 
Particle Image Velocimetry (PIV), 
and Lagrangian Particle Tracking (LPT).  
The latter option will be implemented 
on the linear traverse mentioned above.  

The hardware we employed to acquire the data presented here 
was a Dantec StreamLine hot wire anemometry system.  
We used two kinds of Dantec wires, 
one with 2.5\,$\mu$m diameter and 450\,$\mu$m length, 
and one with 5\,$\mu$m diameter and 1\,mm length.  
We also used the new NSTAP probes developed at Princeton,\cite{Bailey2010,vallikivi:2011} 
which were either 30\,$\mu$m or 60\,$\mu$m long 
%changes due to reviewer comment 6
with a thickness perpendicular to the flow of 100\,nm, 
and a width in the direction of the flow of 2\,$\mu$m.  
Despite their nontraditional shape, 
end conduction effects due to the prongs are probably negligible.\cite{Bailey2010,Hultmark2011}  
The measured spectra of velocity fluctuations of classical probes and NSTAP largely agree, 
with the latter providing increased spatial and temporal resolution.\cite{vallikivi:2011}  
%end of changes due to reviewer comment 6
The probes were calibrated \textit{in situ} against Prandtl (or Pitot-static) tubes 
while varying the fan speed.  
We also used X-probes with 2.5\,$\mu$m wires to measure the Reynolds stresses.  
The angular responses of the X-wires were calibrated with the Dantec calibrator 
using air at standard temperature and pressure.  
The signals were filtered at 30\,kHz 
and sampled at 60\,kHz with a digital acquisition card.

\section{Test Experiments}
\label{sec:testexp}

We measured the characteristics of the turbulence 
at a fixed distance, 7.1\,m, 
from the 186.6\,mm classical grid, 
and on a array of 150 points covering a 60\,cm by 60\,cm square 
centered in the cross section of the tunnel.  
The measurements were made in both air and SF$_6$ 
at different pressures between 
$\unit[1]{bar}$ and $\unit[15]{bar}$.  
The profiles of the flow were measured by Dantec hot wires, 
and the spectra by NSTAPs.  

%new figure due to comment 9
\begin{figure}
\begin{center}
\includegraphics[width=5.5in]{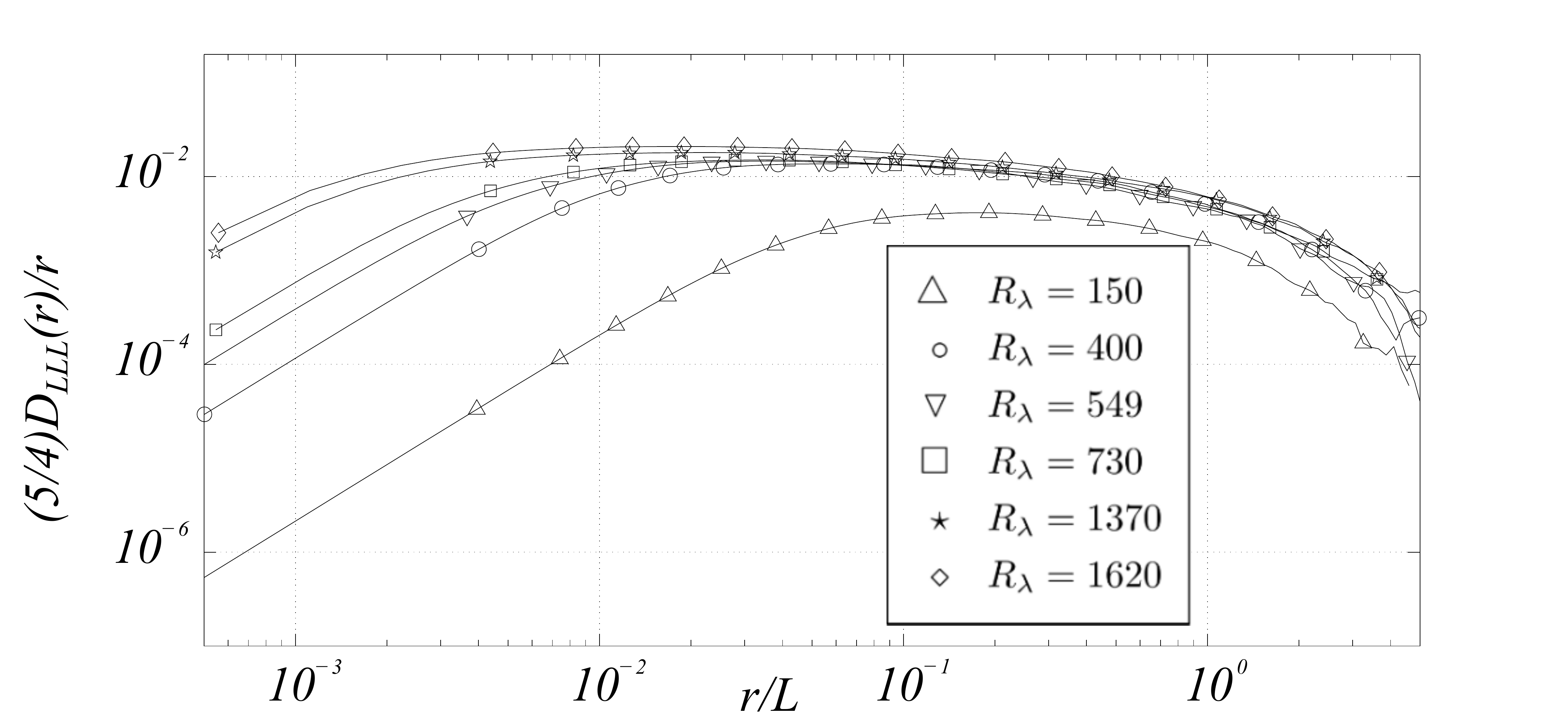}
\caption{We used the peak value of these curves 
as a measure of the energy dissipation rate of the turbulence, 
invoking Kolmogorov's relation $\epsilon = -(5/4) D_{LLL}(r) / r$ 
for the inertial range, 
where $D_{LLL}$ is the third-order structure function.  
}
\label{fig:S3}
\end{center}
\end{figure}
%end changes due to comment 9

Table~\ref{tab:measurement} summarizes the properties of the turbulence.  
These properties were calculated as follows.  
The energy dissipation rate was extracted from the third-order structure functions 
using the 4/5$^{ths}$ law, $D_{LLL}(r) \approx -\frac{4}{5} \epsilon r$ 
in the inertial range, 
so that our practical definition of it was $\epsilon = max(-\frac{5}{4} D_{LLL}(r) / r)$.  
Here $D_{LLL}(r) = \langle \delta u^3(x, r)  \rangle_x$ 
is the third moment of the longitudinal velocity differences, 
$\langle \cdot \rangle_x$ is the average over $x$, 
$\delta u(x, r) = u(x+r) - u(x)$ are the velocity differences, 
$u$, $x$, and $r$ are parallel, 
and we used Taylor's hypothesis to extract $x$ and $r$ from the time series of each probe.  
% new: 
Graphs of the third-order structure functions are shown in Fig.~\ref{fig:S3}.  
% end new.  
We are aware that 
this measure of the dissipation rate is 
smaller than the actual dissipation rate at low Reynolds numbers 
where the inertial range is not well-developed.  
The measure becomes more accurate as the Reynolds number increases.\cite{antonia:2006} 
This has the effect, among others, of slightly inflating the value of the lower Reynolds numbers.  

The amplitude of the velocity fluctuations is $u^\prime = \langle u^2 \rangle_x^{1/2}$.  
The integral scale, $L$, is the integral of the longitudinal correlation functions, 
$L = u^{\prime-2} \int_0^\infty \langle u(x+r)u(x) \rangle_x \, dr$, 
and we used exponential extrapolations to extend the integrals to infinity.\cite{bewley:2012} 
The Taylor scale was evaluated through the isotropic relation, 
$\lambda = (15\nu \langle u^2 \rangle_x / \epsilon )^{1/2}$.  
Since the turbulence was approximately isotropic at all scales,\cite{bewley:2012} 
the Reynolds number is given by $R_\lambda = u^\prime \lambda / \nu$.  
The Kolmogorov scales are given by 
$\eta = (\nu^3 / \epsilon)^{1/4}$, 
and 
$\tau_\eta = \sqrt{\nu / \epsilon}$, 
as usual.

\begin{table}
\caption{The flow parameters for the passive grid experiments.  
$P$ is the pressure of the gas in the tunnel, 
$\rho$ and $\nu$ are the density and viscosity of the gas, respectively, 
$U$ is the mean speed of the flow, 
$u^\prime/U$ is the turbulence intensity, 
$L$ is the streamwise longitudinal integral scale, 
$\epsilon$ is the dissipation rate per unit mass, 
$\eta$ and $\tau_\eta$ are the Kolmogorov length and time scales, respectively, 
$\lambda$ is the Taylor scale, 
and $R_\lambda$ is the Taylor Reynolds number.  
}
\label{tab:measurement}
\begin{tabular}{l|cccccc}
fluid                                         & Air      & SF$_6$ & SF$_6$ & SF$_6$ & SF$_6$ & SF$_6$ \\
$P$ $\unit{[bar]}$                    &1.0      &1.1          & 2.1        & 4.0         & 12         & 15 \\
$\rho$ $\unit{[kg/m^3]}$          & 1.20  & 6.52        & 12.6      & 25.1      & 85.3      & 112 \\
$\nu$ $\unit{[mm^2/s]}$          & 15.2   & 2.32       & 1.21     & 0.602     & 0.184    & 0.143 \\
$U$ $\unit{[m/s]}$                   & 4.31   & 4.16       & 4.11      & 4.11       & 4.08      & 4.08 \\
%RMS velocity $u^\prime$ $\unit{[m/s]}$ & &&&&&\\
$u^\prime/U$ $\unit{[\%]}$       & 2.35  &  3.33      & 3.38      & 3.28       & 3.50      & 3.70 \\
$L$ $\unit{[mm]}$                    & 148   & 139        & 151       & 125        & 125       & 123 \\
$\epsilon$ $\unit{[cm^2/s^3]}$ & 66.0  & 163        & 166       & 159        & 181       & 203 \\
%dissipation rate $\epsilon$ $\unit{[m^2/s^3]}$&0.0012&0.0015&0.0015&0.0015&0.0015&0.0014\\
$\epsilon L/u^3$                      & 0.58  & 0.71       & 0.83       & 0.77      & 0.77      & 0.75 \\
$\eta$ $\unit{[\tcmu m]}$         & 930   & 170        & 100        & 61         & 24         & 19 \\
$\tau_\eta$ $\unit{[ms]}$         & 61     & 13          & 9.1         & 6.3        & 3.2        & 2.6 \\
$\lambda$ $\unit{[mm]}$         & 22.5  & 6.71        & 4.77      & 3.26      & 1.76      & 1.53 \\
$Re_{WT} \times 10^{-3}$       & 43    & 260         & 510        & 1000     & 3400     & 4400 \\
$R_\lambda$                           & 150   & 400         & 549       & 730       & 1370     & 1620 \\
\end{tabular}
\end{table}

Observe in Table~\ref{tab:measurement} 
that changing the Taylor Reynolds number by an order of magnitude 
had a small effect on the integral measures of the flow, 
$u^\prime$ and $L$.  
The effect on the dissipation rate was also small.  
The main effect of Reynolds number variation 
was on the scale at which dissipation occurs.  
%a point that we revisit below.  
%changes for comment 4: 
These observations are in general agreement 
with those reported in the literature,\cite{comte-bellot:1971,Kurian2009,krogstad:2010} 
and point to a main advantage of the facility: 
that the Reynolds number can be varied while fixing the conditions at the large scale.  
%end of changes due to reviewer comment 4

In the sections that follow 
we describe the statistics of the flow at positions midway between the side walls, 
and at various distances, $z/H$, from the floor of the tunnel, 
where $z$ is the height of the probe above the floor 
and $H$ is the height of the tunnel.  
Horizontal cuts, made across the width of the tunnel, 
were at least as good, or better, than the vertical cuts we present, 
in the sense that the deviations from homogeneity across the width were smaller.  
We then discuss the properties of the spectrum of the velocity fluctuations.  
Some analysis of these data
%namely of conditional structure functions, 
% deleted due to comment 5
has already been published in Blum et al.\cite{blum:2011}  

\begin{figure}
\begin{center}
\includegraphics[width=6.5in]{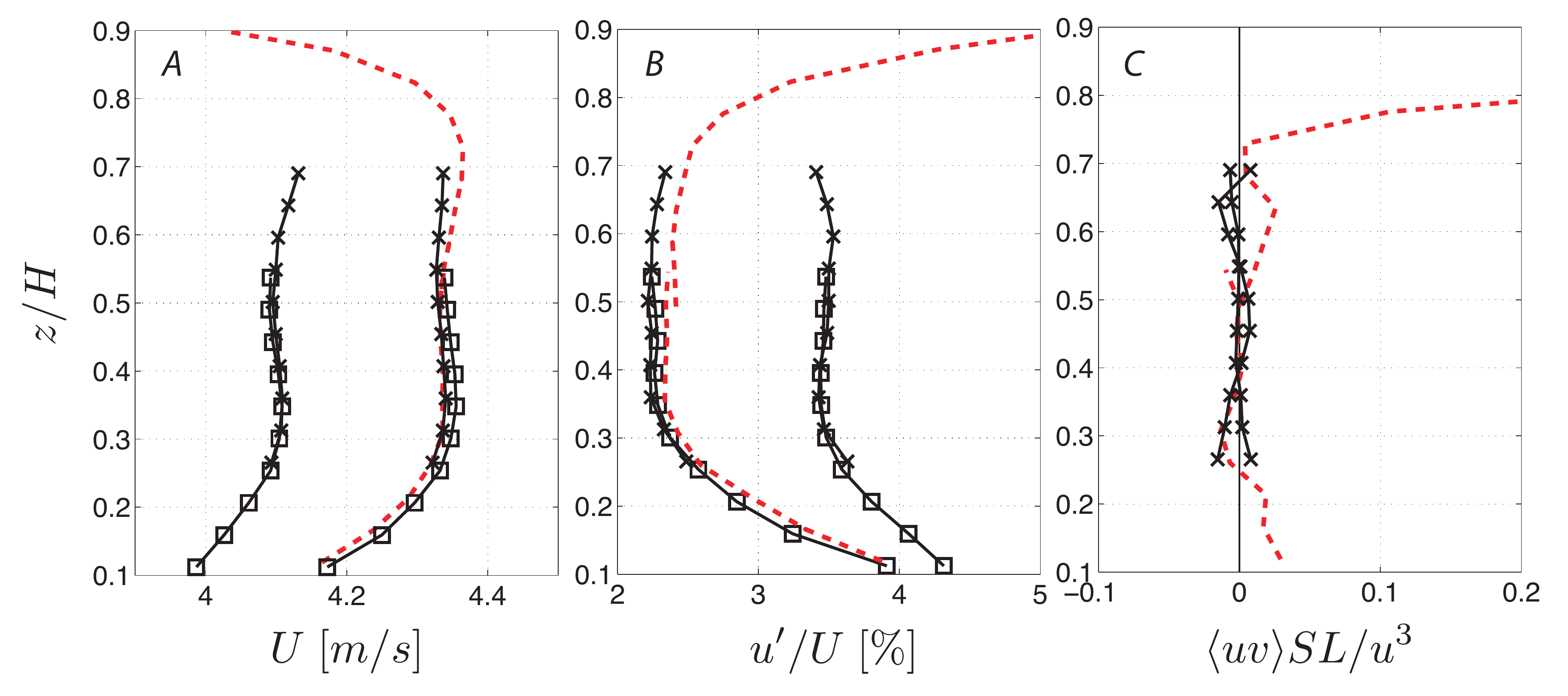}
\caption{
A.  Profile of the mean velocity as a function of vertical distance from the floor of the tunnel.  
The vertical position $z$ is normalized by $H$, 
the distance from the floor to the ceiling.  
The $\times$ symbols were acquired with an X-wire, 
and the $\Box$ symbols with a single wire.  
The dashed red curve was acquired separately in air with X-wires 
mounted on a traverse that made a wider range in $z$ accessible.  
B.  Profile of the turbulence intensity as a function of vertical distance from the floor of the tunnel.  
C.  Profile of the turbulence production by shear as a fraction of the dissipation rate.  
For all of these measurements, the rotation rate of the fan was 20\,Hz, 
while its maximum is 24.5\,Hz 
so that mean flow speeds up to about 5.3\,m/s are possible.  
}
\label{fig:profiles}
\end{center}
\end{figure}

\subsection{Profile of Mean Velocity}

Fig.\ \ref{fig:profiles}A shows the profile of the mean velocity, $U$, % = \langle u \rangle_t$, 
as a function of vertical distance from the floor of the tunnel.  
The shape of the profile was independent 
of the viscosity of the gas.  
The mean velocity was approximately constant across the middle half of the section, 
though there were boundary layers at the top and bottom of tunnel 
as well as a small velocity deficit near $z/H=0.5$.  
As mentioned in Section~\ref{sec:grids}, 
the width of the boundary layers may be related to the way the edge 
of the grid meets the wall of the tunnel.  %~\cite{bewley:2006}.  
The boundary layers were slightly thinner when we used grids with smaller mesh spacing.  
Despite these features, the mean flow is reasonably uniform between 
$z/H$ values of 0.25 and 0.75.  
The mean flow was slower at higher pressures, 
which may indicate that the pump was less efficient there 
than at lower pressures.  

%\begin{figure}
%\begin{center}
%\includegraphics[width=3in]{fromGreg/meanflow_13_2_22.eps}
%\caption{Profile of the mean velocity as a function of vertical distance from the floor of the tunnel.  
%The vertical position $z$ is normalized by $H$, 
%the distance from the floor to the ceiling.  
%The $\times$ symbols were acquired with an X-wire, 
%and the $\Box$ symbols with a single wire.  
%The solid curve was acquired separately in air with X-wires 
%mounted on a traverse that made a wider range in $z$ accessible.  
%The fan speed was 20\,hz.  maximum speed is 24.5\,hz.  
%}
%\label{fig:vertprofile}
%\end{center}
%\end{figure}

\subsection{Profile of Turbulence Intensity}

Fig.\ \ref{fig:profiles}B shows the profile of the turbulence intensity 
$u^\prime/U$ in $\unit{\%}$.  
%where $u^\prime$ is the root-mean-square of the velocity 
%(symbols as before).  
As with the mean flow, the turbulence intensity is approximately constant 
between $z/H$ values of 0.25 and 0.75.  
The anisotropy in the fluctuations, $u^{\prime}/v^{\prime}$, 
where $u^{\prime}$ was in the streamwise direction 
and $v^{\prime}$ was in the spanwise direction, 
was between 1 and 1.1, 
so that the fluctuations were slightly stronger in the streamwise direction, 
as has been observed in grid turbulence before.\cite{comte-bellot:1966}  
The turbulence intensity increased with Reynolds numbers 
for reasons that we do not know, 
though the same phenomenon was observed 
also in the HDG at the DLR in G\"ottingen (described above).  
We speculate that the turbulence decays more slowly at high Reynolds numbers, 
and are now performing experiments to test this idea.  

%\begin{figure}
%\begin{center}
%\includegraphics[width=3in]{fromGreg/intensity_13_2_22.eps}
%\caption{Profile of the turbulence intensity as a function of vertical distance from the floor of the tunnel.  
%Symbols as in Fig.~\ref{fig:vertprofile}.  }
%\label{fig:vertturb}
%\end{center}
%\end{figure}

\subsection{Profile of Turbulence Production by Shear}

Fig.\ \ref{fig:profiles}C shows the profile of the turbulence production by shear, 
$\langle u v\rangle S L / u^{\prime 3}$, 
as a fraction of the energy dissipation rate.  
Here, $u$ and $v$ are the velocity fluctuations in the streamwise and spanwise directions, respectively, 
$\langle u v \rangle$ is the Reynolds stress, 
$S = \Delta U / \Delta z$ is the mean shear rate 
and is derived by taking finite differences of the data in Fig.\ \ref{fig:profiles}A, 
and $u^{\prime 3} / L$ is a measure of the energy dissipation rate per unit mass, $\epsilon$.  
The production of turbulence by shear is negligible, 
being of the order of about 1\% of the turbulence dissipation rate 
within the center part of the flow.  
Its importance increases as expected in the top and bottom boundary layers.  

%\begin{figure}
%\begin{center}
%\includegraphics[width=3in]{fromGreg/production_13_2_27.eps}
%\caption{Profile of the turbulence production by shear as a fraction of the dissipation rate.  
%Symbols as in Fig.~\ref{fig:vertprofile}.  }
%\label{fig:vertprod}
%\end{center}
%\end{figure}

\subsection{Reynolds-stress Anisotropy}

\begin{figure}
\begin{center}
\includegraphics[width=0.5\textwidth]{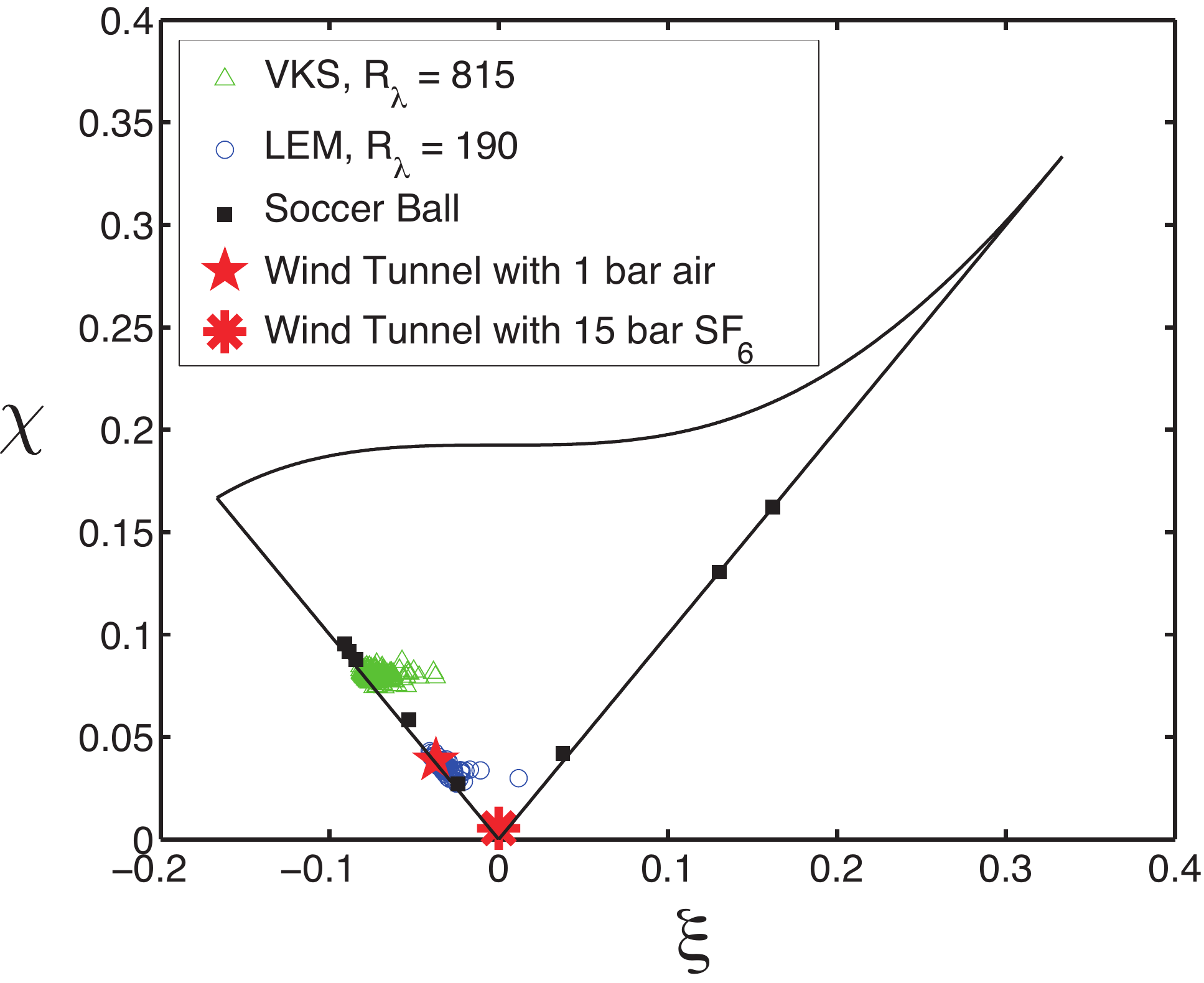}
\caption{Comparing Reynolds-stress anisotropy on the Lumley triangle.  
Here $\xi$ and $\chi$ are two independent invariants of the Reynolds-stress anisotropy tensor 
$b_{ij} = \langle u_i u_j \rangle - (1/3) \langle u_k u_k \rangle \delta_{ij}$, 
defined as $\xi = (b_{ij} b_{jk} b_{ki}/6)^{1/3}$ and $\chi = (b_{ij} b_{ji}/2)^{1/2}$.  
Turbulence in the VDTT was close to isotropic.  
The anisotropy of the turbulence in the soccer ball was adjustable, 
and each of the black squares corresponds to a different setting of the anisotropy.  
In the cases of the VKS and LEM, the cloud of points corresponds to measurements 
made at different positions near the centers of each apparatus.  
}
\label{fig:lumleytriangle}
\end{center}
\end{figure}

To quantify the the Reynolds-stress anisotropy, 
we plot the two invariants of the anisotropic Reynolds stress tensor 
$\langle u_i u_j \rangle - \frac{1}{3} \langle u_k u_k \rangle \delta_{ij}$ 
on the so-called Lumley triangle.\cite{lumley1978,pope2000}  
We assume that the two transverse components are statistically the same, 
\textit{i.e.}, $v^{\prime} = w^{\prime}$, 
%$\langle v^2 \rangle = \langle w^2\rangle$, $\langle u v \rangle = \langle u w \rangle$, 
and that $\langle vw \rangle = 0$.  
We compare the Reynolds-stress anisotropy of the turbulence in the VDTT with other laboratory flows, 
including: 
the von K\'arm\'an swirling flow between counter-rotating disks 
(``the French washing machine''),\cite{voth2002} 
the Lagrangian Exploration Module (LEM),\cite{zimmermann2010} 
and the ``soccer-ball''.\cite{chang2012}  
The data for the French washing machine and the LEM 
were obtained using three-dimensional particle tracking,\cite{ouellette:2006} 
while the data for the soccer-ball were measured with laser-Doppler velocimetry 
and similar assumptions on the two transverse velocity components 
were used to construct the Reynolds stress tensor.  
As shown in Fig.~\ref{fig:lumleytriangle}, 
the turbulence produced in the VDTT was as close to isotropic 
as any other flow, 
% slight adjustment here: 
and more isotropic at higher Reynolds number.  
%The degree of isotropy increased with Reynolds number.  

\subsection{Turbulence Spectra}

\begin{figure}
\begin{center}
\includegraphics[width=0.8\textwidth]{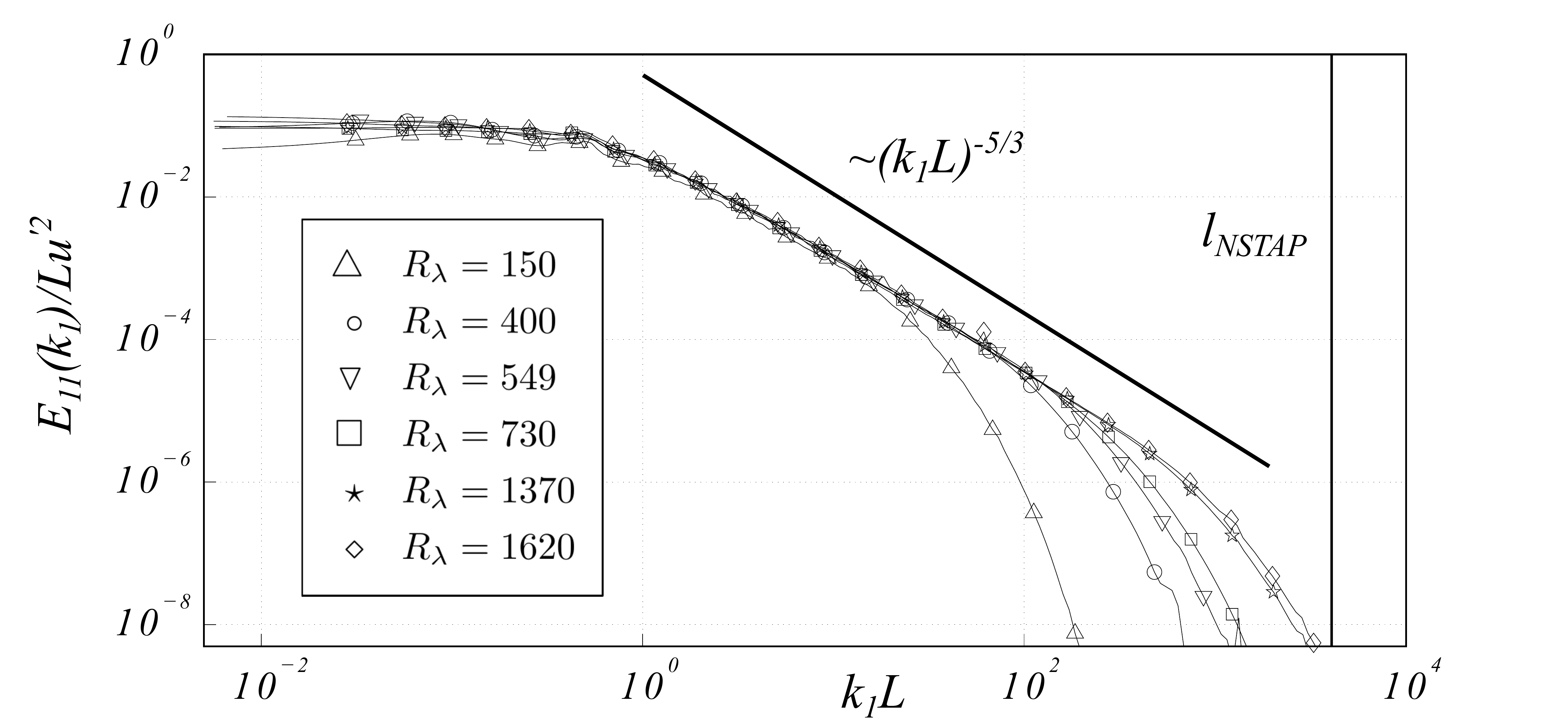}
\caption{The extension of the inertial range toward smaller scales with increasing Reynolds number.  
The curves correspond to the data summarized in table~\ref{tab:measurement}, 
% changed in response to reviewer: 
with 
%with the Reynolds number increasing from left to right as one follows 
%the intersections of the curves with abscissa.  
%As seen in the table, 
the velocity and length scales 
being approximately $u^\prime \approx 0.14\,m/s$ and $L \approx 130\,mm$, respectively, 
%were approximately the same 
for all experiments.  
The vertical bar, marked $l_{NSTAP}$, corresponds to the size of the probe, 
which was 30\,$\mu$m.  
% new: 
Compared to the Kolmogorov scale, the probe was 
$\eta / l_{NSTAP} = $ 31, 5.7, 3.3, 2.0, 0.80, and 0.63 
times smaller in order of increasing Reynolds number.  
%which supports the view that the large wavenumber spectral cutoff 
%was probably the result of viscous dissipation in the turbulence 
%and not due to filtering by the probe.  
% end new.  
}
\label{fig:spectra}
\end{center}
\end{figure}

Figure~\ref{fig:spectra} shows the development of the inertial range in the energy spectra 
through extension of the small scales.  
These are 30\,micron NSTAP data in air and 1, 2, 4, 12 and 15\,bar SF$_6$, 
and the measurements were made in the center of the tunnel where 
the flow was approximately homogeneous.  
The spectra, 
$E_{11}(k_1) = \int\int_{-\infty}^{\infty}\frac{E\left(k\right)}{2\pi k^2}\left(1-\frac{k_1^2}{k^2}\right)\mathrm{d}k_2\mathrm{d}k_3$, 
are normalized by the integral quantities, $u^\prime$ and $L$, 
which change only a little with Reynolds number, 
as can be seen in Table~\ref{tab:measurement}.  
The collapse of the spectra at large scales supports the view 
that the dynamics at these scales 
are Reynolds number independent.  
In other words, they are set 
by the boundary conditions 
and are not strongly influenced by the material properties of the gas.  
This demonstrates that with the VDTT 
we control the large scales 
and modulate only the small scales by changing the pressure.  
%changes due to comment 8
As is well known, 
the spectra do not scale exactly as $k^{-5/3}$.  
Indeed, the spectra have a Reynolds-number dependent structure, 
a point we revisit below after first comparing the spectra acquired with different probes.  
%end changes

\begin{figure}
\begin{center}
\includegraphics[width=4.2in]{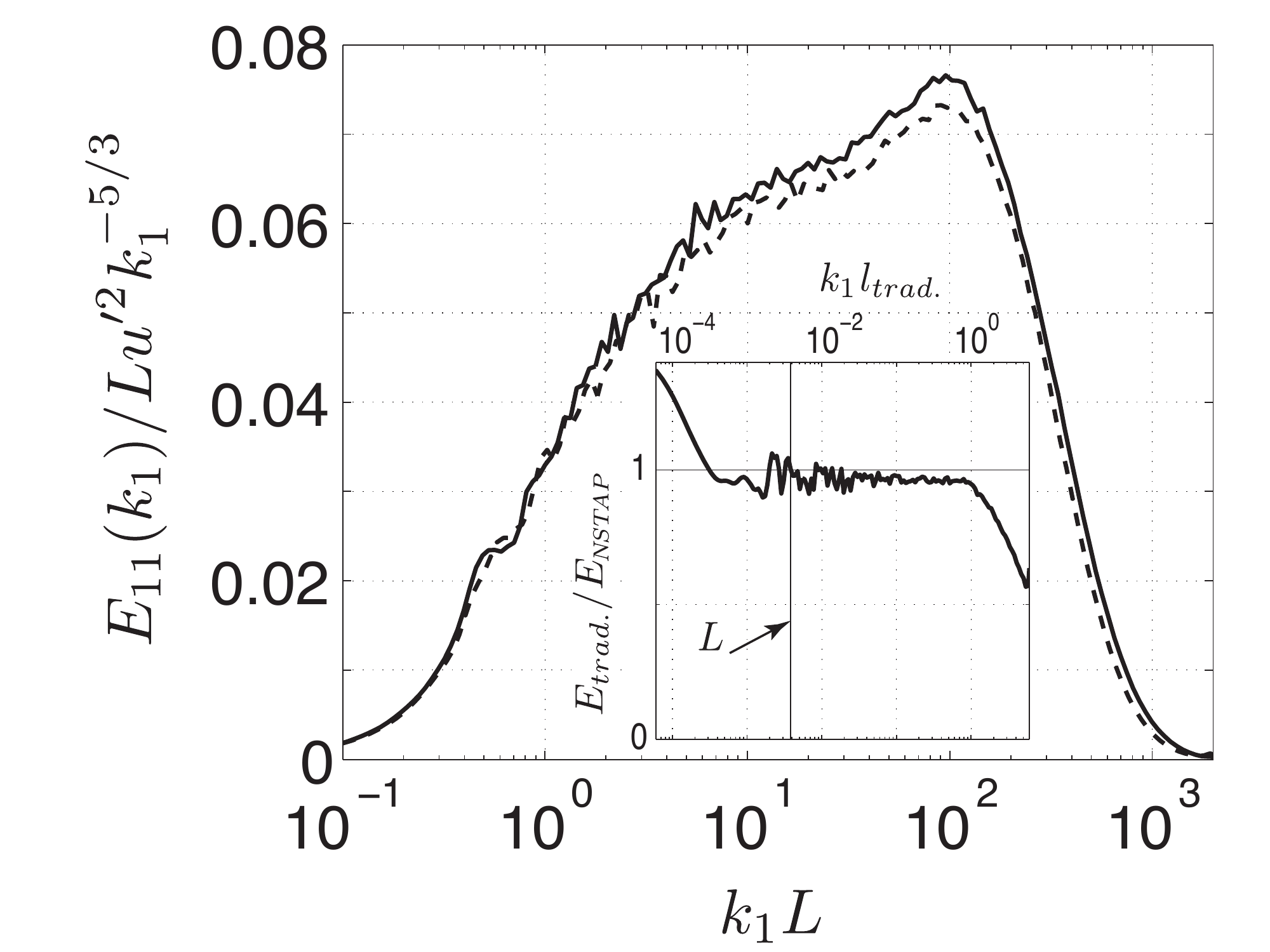}
\caption{A comparison of a spectrum acquired with an NSTAP ($E_{NSTAP}$, solid) 
to one acquired with a standard Dantec hot wire ($E_{trad.}$, dashed).  
These data were acquired at 4\,bar, so that the Taylor Reynolds number was about 730.  
The inset shows the ratio of the two spectra, 
which is approximately constant 
throughout the inertial range, 
and down to the scale that corresponds to the size of the Dantec hot wire probe, 
$l_{trad.}$ = 450\,$\mu$m.  
At smaller scales, the Dantec wire was less sensitive than the NSTAP.  
The vertical bar corresponds to the integral scale, $L$.  
}
\label{fig:spectracomparison}
\end{center}
\end{figure}

Figure~\ref{fig:spectracomparison} 
compares a spectrum acquired with a Dantec probe to one with an NSTAP 
at the same moderate Reynolds number 
(it is the 4\,bar data in Table~\ref{tab:measurement}).  
The key point is that the shapes of the two spectra are nearly identical, 
down to the scale of the Dantec probe, $l_{trad.}$, 
which builds confidence in the NSTAP data.  
The deviation of their ratio from one at the largest scales, visible in the inset, 
is probably due to the usual problems 
with convergence at these scales, which are much longer than the integral scale.  
At scales smaller than the Dantec probe, or at large $k_1 l_{trad.}$, the ratio falls off.  
This roll-off probably indicates that the Dantec probe lost sensitivity relative to the NSTAP 
for $k_1 l_{trad.} > 1$.  
In addition to this spatial filtering effect, 
there are hints that temporal filtering of the hot-wire system 
can influence the measurement at high frequencies.\cite{Ashok2012}  
%The most recent studies show that 
%the frequency response of different anemometer and probe combinations 
%can vary substantially.\cite{Hutchins2013}  % this paper still hasn't been published.  
This ongoing work will need further careful consideration.  

\begin{figure}
\begin{center}
\includegraphics[width=4.2in]{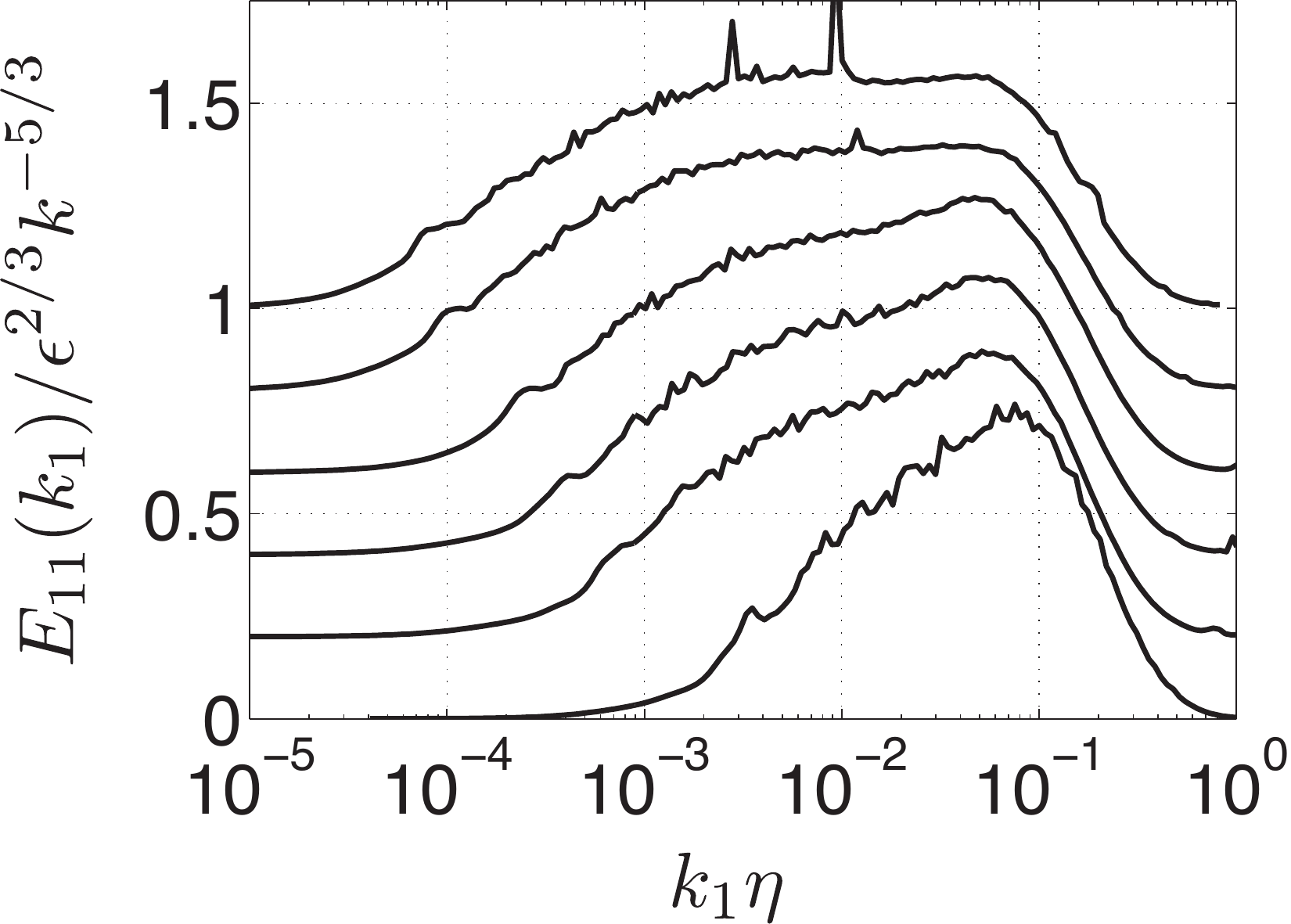}
\caption{Compensated spectra.  
As in Fig.~\ref{fig:spectra}, the data correspond to those summarized in Table~\ref{tab:measurement}.  
%The Reynolds number increases from bottom to top, 
%with each curve shifted by 0.2 with respect to the one below it (except for the bottom one).  
The spectra are shifted vertically by 0.2 in order to set them apart, 
with the lowest Reynolds number data at the bottom.  
The extension of the inertial range is visible, 
as is the flattening of the spectra with increasing Reynolds number.  
}
\label{fig:compensatedspectra}
\end{center}
\end{figure}

Figure\ \ref{fig:compensatedspectra} shows the same spectra as in Fig.\ \ref{fig:spectra}, 
but in the Kolmogorov variables.  
The inertial range is the approximate plateau in the curves.  
% new sentence: 
Both the general structure of the curves and their development with Reynolds number 
are similar to what can be found in the literature.\cite{Saddoughi1994,mydlarski:1996,Donzis2010}  
The so-called bottleneck, or the bump on the right side of the inertial range, 
initially grows more pronounced with increasing Reynolds number
before losing prominence at the highest Reynolds numbers.  
%new/adjusted: 
Overall, the spectra become more horizontal in the inertial range 
%The overall slope of the spectra in the inertial range grows more shallow 
with increasing Reynolds number, 
which can be interpreted as a slow approach to $k^{-5/3}$ scaling.\cite{mydlarski:1996}  
% end new.  
% Detailed interpretation will require more attention to probe effects, 
% like filtering by the electronics, probe size effects, prong effects, temperature drifts, etc.  

\subsection{Scale Separation}

\begin{figure}
\begin{center}
\includegraphics[width=2.5in]{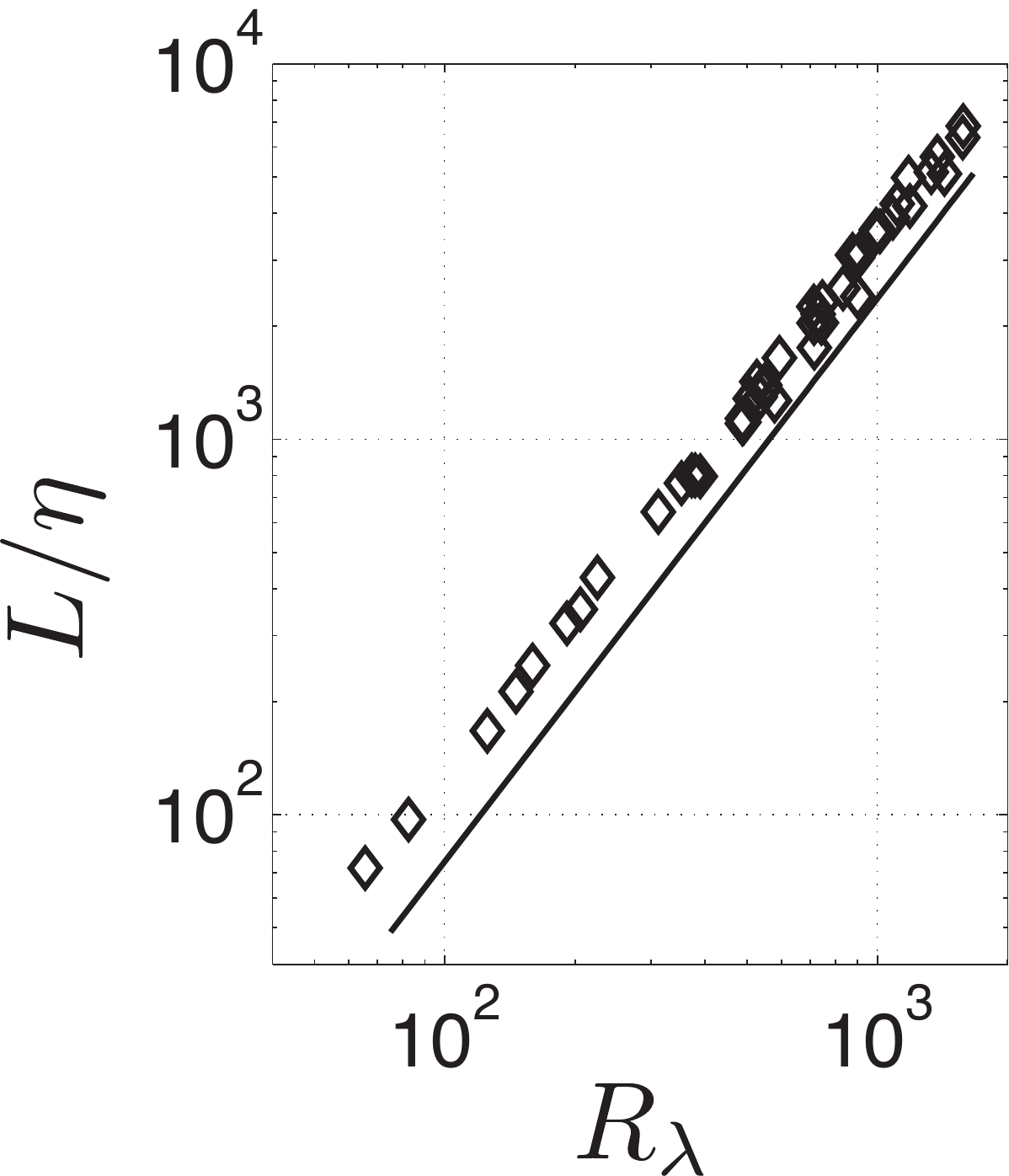}
\caption{Scale separation arises as the Reynolds number increases.  
Here we include data acquired at other pressures and fan speeds 
than included in the previous graphs or Table~\ref{tab:measurement}.  
The solid line is a 3/2 power law, as discussed in the text.  
}
\label{fig:scaleseparation}
\end{center}
\end{figure}

Figure\ \ref{fig:scaleseparation} shows the ratio of the integral scale, $L$, 
to the Kolmogorov scale, $\eta$.  % = (\nu^3/\epsilon)^{1/4}$.  
The integral scale is a measure of the size of the sweeping, energy containing motions, 
while the Kolmogorov scale measures the scale of the sharp, dissipative gradients.  
As mentioned in the introduction, 
higher Reynolds numbers are related to increased separation between these large and small scales.  
This expression can be made more precise.  
Using the relation $\epsilon \approx u^{\prime 3}/L$, 
it is easy to show that $L/\eta \sim R_\lambda^{3/2}$.  
The line in the figure is a 3/2 power law of arbitrary amplitude, 
and it follows the data reasonably well.  
This illustrates one of the chief advantages of the tunnel, 
which is that scale separation can be achieved by changing the pressure in the tunnel, 
and so by modulating the small scales alone.  
In this way, we separate the influence of scale separation on the small-scale dynamics 
from the influence of changes in the large-scale structure of the flow.  

\section{Conclusion and Outlook}

The new high-pressure, high-turbulence wind tunnel in G\"ottingen, 
the VDTT, 
makes experimental measurement of the structure and dynamics 
of nearly homogeneous and shearless turbulence 
possible at higher Reynolds numbers than before. 
With passive grids we reach Taylor Reynolds numbers of 1600, 
whereas comparable studies reach about 870 (with an active grid).\cite{shen:2002}  
To characterize the quality of the flow, 
hot-wire measurements were made behind a classical grid turbulence generator.  
Using tools that we describe in the paper, 
we will in the future reach even higher Reynolds numbers 
and introduce new Lagrangian measurement techniques.  

With the addition of the active grid in the VDTT, 
we expect to reach Reynolds numbers 
before attainable only in the atmospheric boundary layer.\cite{pearson:2001}  
That is, 
we will produce steady homogeneous and isotropic conditions, 
whereas existing data were acquired in unsteady inhomogeneous and anisotropic flows.  
As can be seen in Table~\ref{tab:prediction}, 
very high Reynolds numbers up to 
$R_\lambda$ at least 4200 will be possible with the active grid.  
These estimate is based on our initial experience with the new active grid 
in an open-circuit air tunnel, 
which we will publish separately, 
and where measurements were made 
at the downstream end of the tunnel.  
The Reynolds number will be higher at the upstream ends of the test sections, 
though the decay of Reynolds number with distance from the grid is typically slow.\cite{krogstad:2010}  
Further optimization of the active grid may yield yet higher Reynolds numbers, 
as has been observed with other active grids.\cite{poorte:2002}  
In this way, we may reach Reynolds numbers up to 8000.  

\begin{table}
\caption{Parameters predicted for the 
turbulence created by an active grid 
in the VDTT.  
The first line contains preliminary measurements made in the separate wind tunnel 
at a distance $d$ = 9\,m downstream of the active grid designed for the VDTT.  
The next lines are extrapolations based on these measurements to conditions in the VDTT.  
First, we assume that the turbulence intensity and dissipation rate 
will remain constant, 
and in the last line, 
we make an educated guess 
of what the VDTT might produce at the upstream end of a test section 
after optimization of the active grid.  
}
\label{tab:prediction}
\begin{tabular}{cccc|ccccccc}
Working & Pressure & $\rho$ & $\nu$ & $U$ & $d$ & $u^\prime/U$ & $\epsilon$ & $R_\lambda$ & $\eta$ & $\tau_\eta$ \\
fluid & $\unit{[bar]}$ & $\unit{[kg/m^3]}$ & $\unit[10^{-6}]{[m^2/s]}$ & $\unit{[m/s]}$ & $\unit{[m]}$ & $\unit{[\%]}$ & $\unit{[m^2/s^3]}$ & & $\unit{[\tcmu m]}$ & $\unit{[ms]}$ \\
\hline
Air          & 1   & 1.29  & 14.0  &  11   &  9  &  6    &  0.46    &  620    &  280  &  5.5  \\
\hline
Air          & 1   & 1.29  & 14.0  &  5.0  &  9  &  6    &  0.045  &  430    &  500  &  18   \\
SF$_6$  & 1   & 5.86  & 2.64  &  5.0  &  9  &  6    &  0.045  &  1000  &  140  &  7.7  \\
SF$_6$  & 15 & 107  &  0.15  &  5.0  &  9  &  6    &  0.045  &  4200  &  17    &  1.8  \\
SF$_6$  & 15 & 107  &  0.15  &  5.0  &  2  &  20  &  1.5      &  8000  &  6.9   &  0.32
\end{tabular}
\end{table}

With the addition of the linear motor and camera system, 
the Lagrangian properties of the turbulence in the VDTT will become accessible.  
The Lagrangian approach coupled with 
conventional Eulerian measurements under well-controlled conditions 
will provide a new perspective on fundamental turbulence questions.  
With its special properties the VDTT 
will make possible experiments in a 
well-understood and well-controlled flow at the highest turbulence levels 
yet possible in the laboratory, 
and with measurable spatial and temporal scales of motion.  
Therefore, it will make it possible to address problems 
important to environmental, atmospheric, and ocean 
sciences, to engineering and astrophysics.

\begin{acknowledgments}

The  establishment of the  G\"ottingen turbulence facility with the VDTT  started in 2003 when Eberhard Bodenschatz (EB) was offered a position at the Max Planck Institute for Flow Research (now Max Planck Institute for Dynamics and Self-Organization, MPIDS) and had received, in addition to his generous startup by the Max Planck Society, additionally 1 Mio Euro from the Volkswagen Foundation.  EB realized that a high Reynolds number turbulent fluid flow  with experimentally resolvable scales was needed to measure Lagrangian  and Eulerian properties in turbulence under controlled 
conditions. % +s
Based on his prior experience in Rayleigh-B{\'e}nard convection in compressed gases \cite{debruyn:1996} he decided that the best approach would be to use compressed gases combined with a slow speed wind tunnel and an active grid.  Guenter Ahlers  from the University of California at Santa Barbara encouraged him to use the high-density gas SF$_6$, that makes it possible to reach low kinematic viscosity and thus high Reynolds numbers at relatively low pressures. In a leap of faith he decided that it must be possible to operate modern electronics, including high speed cameras in a pressurized environment without any extra protection.  Thus he designed the VDTT with only rudimentary optical access. This, in turn,  allowed  a simple design with an inexpensive pressure vessel. EB was involved in all aspects of the design, manufacturing of the equipment and in the design of the experiments presented here.  This also includes the building of the hall that now holds  the experiments and the gas handling systems. 

The initial study of the properties and requirements of the pressurized flow in the tunnel were prepared by Micka\"el Bourgoin and EB in 2004 with 
%the 
help from Zellman Warhaft at Cornell University.  EB and Siegfried Maier  (SM) coordinated the original design of the VDTT and of the experimental hall with separated foundations and many other features necessary for the 
safe % save->safe
operation of  the VDTT.  In May 2004 TLT Turbo GmbH, Zweibr\"ucken, Germany had finished the first design and made a first offer. Later this year Holger Nobach (HN) took over as project coordinator.  HN, EB, SM with help from Stefan Luther and Haitao Xu solidified the design in long discussions with TLT.  Finally, in June 20, 2006 a final contract for the construction of the tunnel was signed. From then on the project coordination of the building of the tunnel was coordinated by HN and EB.   The SF$_6$ gas handling system was designed, built and installed by  DILO Armaturen und Anlagen GmbH, Germany in collaboration with EB, HN and SM.  

During the design of the VDTT some setbacks were found and 
modifications % +s
needed to be made. For example in  May 2008, after running the tunnel for a first time the heat exchanger lost mechanical integrity due to resonances of the mechanical structure and needed to be fully replaced. Finally in May 7, 2009 the tunnel was inaugurated. Udo Schminke (US) and his colleagues in the mechanical workshop of the 
MPIDS % +this
provided their experience to the construction of the VDTT and its modifications every since. Bernhard Krey and his colleagues from management engineering  participated directly in the installation of the tunnel and in the proper specifications of the safety requirements of the hall.  Very important to mention are also Daniela Wurst and Alfred Schmucker from the building department  of the Max Planck Society who were essential in designing and constructing the hall that houses the facility.  Last but not least we are thankful to the architect Hans-Jochen Schwieger for the very nice and functional design of the building. 

The installation of the tunnel  interior and its operation has  heavily relied and continues to heavily rely on the design engineer Artur Kubitzek (AK)  and his group of technicians.  AK joined the Bodenschatz group in December 2006 and was joined by the electric technician Andreas Renner in December 2007 and by mechanical technician Andreas Kopp in November 2009. We are especially grateful to Andreas Kopp for the excellent technical support in operating this very complicated facility.  

From  January 2008, Greg Bewley (GB) with help from HN, HX ,EB and  AK setup a traditional hotwire measurement system with a passive grid in the tunnel. Starting roughly at the same time GB led the design, construction and test of the active grid that has shown excellent performance in the historical tunnel mentioned above.   Currently a new and final version for the active grid  designed by GB, with help from AK and US is in production and will be installed this fall. We are also grateful for the electronics shop of the MPIDS, in particular to Ortwin Kurre for the installation of the wiring system. 

The  design of the sled system has been under the leadership of HX, who was supported by Hengdong Xi, HN, GB and EB.  Sebastian Lambertz build a prototype of the sled and with HX used it to conduct measurements on inertial particle dynamics.  Currently the lid  and sled system has been designed by iW Maschinenbau GmbH and is to be installed in the fall of this year. 

GB with help from Michael Sinhuber (MS) and EB conditioned the tunnel to produce homogeneous turbulence. GB and MS  conducted the passive grid measurements presented here.  G\"unter Schewe and Helmut Eckelmann helped 
to identify and solve problems with the flow and vibration in the tunnel.  
Visiting student Andrea Costanzo helped diagnose the flow separation 
that occurred in the expansion upstream of the upper test section.  
GB with Greg Voth and  Florent Lachauss\'ee, conducted the measurements of active grid turbulence that we report here.  

Measurements with adequate resolution were made possible through collaboration 
with Margit Vallikivi, Marcus Hultmark, and Alexander Smits at Princeton University.  
They manufactured the NSTAPs that we used, 
and have been a vital resource in their proper use.  
We look forward to further fruitful collaboration with them, 
as we continue to explore turbulence with the VDTT.  

GB led and coordinated the writing of this manuscript to which all authors contributed.. 

Financial support by the Volkswagen Stiftung and the Max Planck Society is gratefully acknowledged .

\end{acknowledgments}

%\bibliography{bibliography,historic,vdtt_sec,activegrid,linearmotor,additionalliterature}
% Run this once to generate your BBL file. Then copy the contents of your BBL file into your main latex file, commenting out "\bibliography"

%

\end{document}